\documentclass[twocolumn,aps,prb]{revtex4-1}
\usepackage{amsmath,amssymb}
\usepackage{graphicx}
\usepackage{epstopdf}
\usepackage[usenames]{color}

\newcommand{\be}{\begin{equation}}
\newcommand{\ee}{\end{equation}}

\newcommand{\bea}{\begin{eqnarray}}
\newcommand{\eea}{\end{eqnarray}}
\newcommand{\ra}{\rangle}
\newcommand{\la}{\langle}
\renewcommand{\tt}{{\tilde{t}}}

\newcommand{\dg}{{\dagger}}
\newcommand{\pdg}{{\phantom\dagger}}
\newcommand{\tphi}{{\tilde\phi}}

\begin{document}

\title{Chiral Mott insulator with staggered loop currents in the fully frustrated Bose Hubbard model}
\author{Arya Dhar$^1$, Tapan Mishra$^{2}$, Maheswar Maji$^3$, R. V. Pai$^4$, Subroto Mukerjee$^{3,5}$ and
Arun Paramekanti$^{2,3,6,7}$}
\affiliation{$^1$ Indian Institute of Astrophysics, Bangalore 560 034, India}
\affiliation{$^2$ International Center for Theoretical Sciences(ICTS),
Bangalore 560 012, India}
\affiliation{$^3$ Department of Physics, Indian Institute of Science, Bangalore 560 012, India}
\affiliation{$^4$ Department of Physics, Goa University, Taleigao Plateau, Goa 403 206, India}
\affiliation{$^5$ Centre for Quantum Information and Quantum Computing, Indian Institute of Science,
Bangalore 560 012, India}
\affiliation{$^6$ Department of Physics, University of Toronto, Toronto, Ontario, Canada M5S 1A7}
\affiliation{$^7$ Canadian Institute for Advanced Research, Toronto, Ontario, M5G 1Z8, Canada}

\begin{abstract}
Motivated by experiments on Josephson junction arrays in a magnetic field
and ultracold interacting atoms in an optical lattice in the
presence of a `synthetic' orbital magnetic fields, we study the ``fully frustrated''
Bose-Hubbard model and quantum XY model with half a flux quantum per lattice plaquette. Using
Monte Carlo simulations and the
density matrix renormalization group method, we show that
these kinetically frustrated boson models admit
{\it three} phases at integer filling: a weakly interacting chiral superfluid phase with staggered loop currents which spontaneously
break time-reversal symmetry, a conventional Mott insulator at strong coupling, and a remarkable ``chiral Mott insulator'' (CMI)
with staggered loop currents sandwiched between them at intermediate correlation. We discuss how the
CMI state may be viewed as an exciton condensate or a vortex supersolid, study a Jastrow variational
wavefunction which captures its correlations,  present results
for the boson momentum distribution across the phase diagram,
and consider various experimental implications of our phase diagram.
Finally, we consider generalizations to a staggered flux Bose-Hubbard model and a two-dimensional (2D)
version of the CMI in weakly coupled ladders.
\end{abstract}

\maketitle

\section{introduction}
The effect of frustration in generating unusual states of matter such as fractional quantum Hall fluids or
quantum spin liquids is an important and recurring theme in the physics of condensed matter systems.
\cite{stone,balents.nature2010}
Recently, research in the field of ultracold atomic gases has begun to explore this area,
spurred on by the creation of artificial gauge fields using
Raman transitions in systems of cold
atoms.\cite{spielman.nature2009,bloch.prl2011}
These gauge fields can be used to thread fluxes through the plaquettes of optical lattices giving rise to ``kinetic frustration''
by producing multiple minima in the band dispersion and frustrating simple Bose condensation into a single non-degenerate
minimum.
Similarly, time-dependent shaking of the optical lattice \cite{Struck.science2011, Struck.prl2012} or populating higher bands of an optical lattice \cite{hemmerich.naturephys2011}
can be used to control the sign of the
hopping amplitude in an optical lattice, again leading to such ``kinetic frustration''.
For bosonic atoms with weak repulsion, such kinetic frustration gets resolved in a manner such that the resulting
superfluid state can have a broken
symmetry corresponding to picking out a particular linear combination of the different minima.
 \cite{hemmerich.naturephys2011,dassarma.prl2008,tosi.prl2005,hemmerich.prl2008,sengupta.epl2010}
Increasing the strength of the interactions at commensurate
filling can be expected to eventually yield a Mott insulator (MI)
with the motion of the bosons quenched, which thus renders the kinetic
frustration ineffective. In a synthetic flux and at strong coupling, the fully gapped MI is
identical to the one expected for the same lattice without a
frustrating flux per plaquette; \cite{fisher.prb1989} this simply means that at strong coupling,
we can adiabatically
remove the flux without encountering a quantum phase transition. However,
there could exist a state intermediate to the superfluid and the MI
described above for which charge motion has been suppressed enough
to open up a gap but not to restore the broken symmetry associated with
frustration. Such a `weak' Mott insulating state, while it is globally incompressible, supports 
significant local boson number fluctuations,
so bosons can `sense' the magnetic flux on a plaquette.
In a recent paper, we have found numerical evidence for
the existence of such a remarkable intermediate state in frustrated two-leg
ladders of bosons for the so-called Fully Frustrated Bose Hubbard (FFBH) model which has
half-a-flux-quantum per plaquette. We call this state a ``chiral Mott insulator''
(CMI) since it is fully gapped due to boson-boson interactions, exactly like an ordinary
Mott insulator, and in addition possesses chiral order associated with the {\it spontaneously} broken
time-reversal symmetry arising from resolving the kinetic frustration. The superfluid state
of this system also possesses this chiral order and we thus dub it a chiral superfluid
(CSF).\cite{dhar.pra2012} Other recent studies have also focussed on a variety of such
exotic states driven by
``ring-exchange'' interactions, \cite{motrunich0,motrunich1,motrunich2,motrunich3,motrunich4,motrunich5}
which again arise in a similar fashion driven by virtual charge fluctuations in a Mott insulator.

In this paper, we discuss further details of our work on this FFBH ladder model, and its
close cousin, the fully frustrated quantum XY model to which it reduces at high filling factors.
We also discuss how one might stabilize such a Mott insulator in higher dimensions by
considering weakly coupled ladders.
Such a CMI may also be viewed as a bosonic Mott insulating version of the
staggered loop current
states \cite{ddw.prb2001,marston.prb2002,troyer.prl2003} studied in the context of pseudogap
physics in the high temperature
cuprate superconductors.

Classical analogues of the CMI and CSF states have been studied in the past.\cite{tosi.prl2005,hemmerich.prl2008,sengupta.epl2010}
 The simplest classical model displaying analogous phases is the fully frustrated
$XY$ model in two dimensions.
\cite{jayaprakash.prb1983,olsson.prl1995} At small but finite
temperature, this model has a phase with algebraic $U(1)$ order for
the spins along with a  staggered pattern of vorticity associated
with each plaquette corresponding to broken $Z_2$ symmetry. This is
the analogue of the CSF phase. As the temperature is increased, the
$U(1)$ symmetry is restored while the $Z_2$ symmetry continues to be
broken in a state that is the analogue of the CMI. Upon further
increasing the temperature, the $Z_2$ order is restored yielding a
completely disordered state, which is the analogue of the
featureless CMI. Analogs of these phases have also been found in another
classical models which will be discussed later~\cite{granato.prb1993,Granato2}.

The CSF phase has been studied in a variety of frustrated quantum
models of bosons but without the strong correlations required to
obtain the CMI
phase.~\cite{orignac_giamarchi,nishiyama,cooper.prb2010,cha.pra2011}
The effect of correlations in conjunction with frustration has been
investigated for models of fermions and quantum
spins.\cite{ddw.prb2001,giamarchi.prl2009,capponi,marston.prl2002,troyer.prl2003,kolezhuk.prl,alhassanieh,scholl.annals,Kolezhuk,mcculloch}
 In both cases, analogs of the CSF state are obtained as staggered current (for fermions) or gapless chiral (for spins) states. An
 analog
 of the CMI state has been found for fermions  in a ladder model.\cite{marston.prb2002} For spins, the analog of the CMI would be a spin gapped
state with vector chiral order, which has been proposed for easy-plane frustrated magnets.\cite{lecheminant.prb2001} Studies on microscopic spin models for a CMI-like
state suggest that it coexists with either dimer order (for half integer spin) or topological order (for integer spin).\cite{hikihara,zarea}

In this paper, we study the Bose-Hubbard ladder of
ref.~\onlinecite{dhar.pra2012} using a  variety of different techniques to
elucidate the nature and properties
 of the CSF and CMI. We employ mean field theory, mapping onto an effective classical model and a variational Monte-Carlo method
 for
this purpose. We also provide physical pictures for the CMI phase as a supersolid of vortices and a condensate of neutral excitons.
The outline of the paper is as follows: In section II, we introduce the microscopic model with a summary of results obtained from
numerics in our previous work. In section III, we perform a mean-field calculation which provides a good description of CSF phase
but is unable to describe the CMI phase. Then in section IV, we write down a rotor model for our Hamiltonian which allows us to map
 it onto a model, which can be studied using classical Monte-Carlo simulations. Numerics on this model have been performed in ref.~\onlinecite{dhar.pra2012}
and show the existence of the CMI. This resulting model can be used for comparison to the classical frustrated models mentioned
earlier. In section V, we provide details of our density matrix renormalization group (DMRG) studies on the
Bose-Hubbard model, which
also reveals a CMI phase. In section VI, we provide physical pictures of the CMI as a supersolid of vortices or a condensate of
neutral excitons while in section VII, we perform a variational Monte-Carlo study of a candidate Jastrow wavefunction for the
CMI state, showing that it correctly captures the essential correlations of the CMI state.
Section VIII provides a short description of experiments that can be performed to detect the CMI state in Josephson
junction arrays or cold atom systems.
Section IX discusses generalizations to a staggered flux model and to higher dimensions, and we conclude in
Section X with a summary.

\section{model and phase diagram}

\begin{figure}[t]
\includegraphics[width=3.0in]{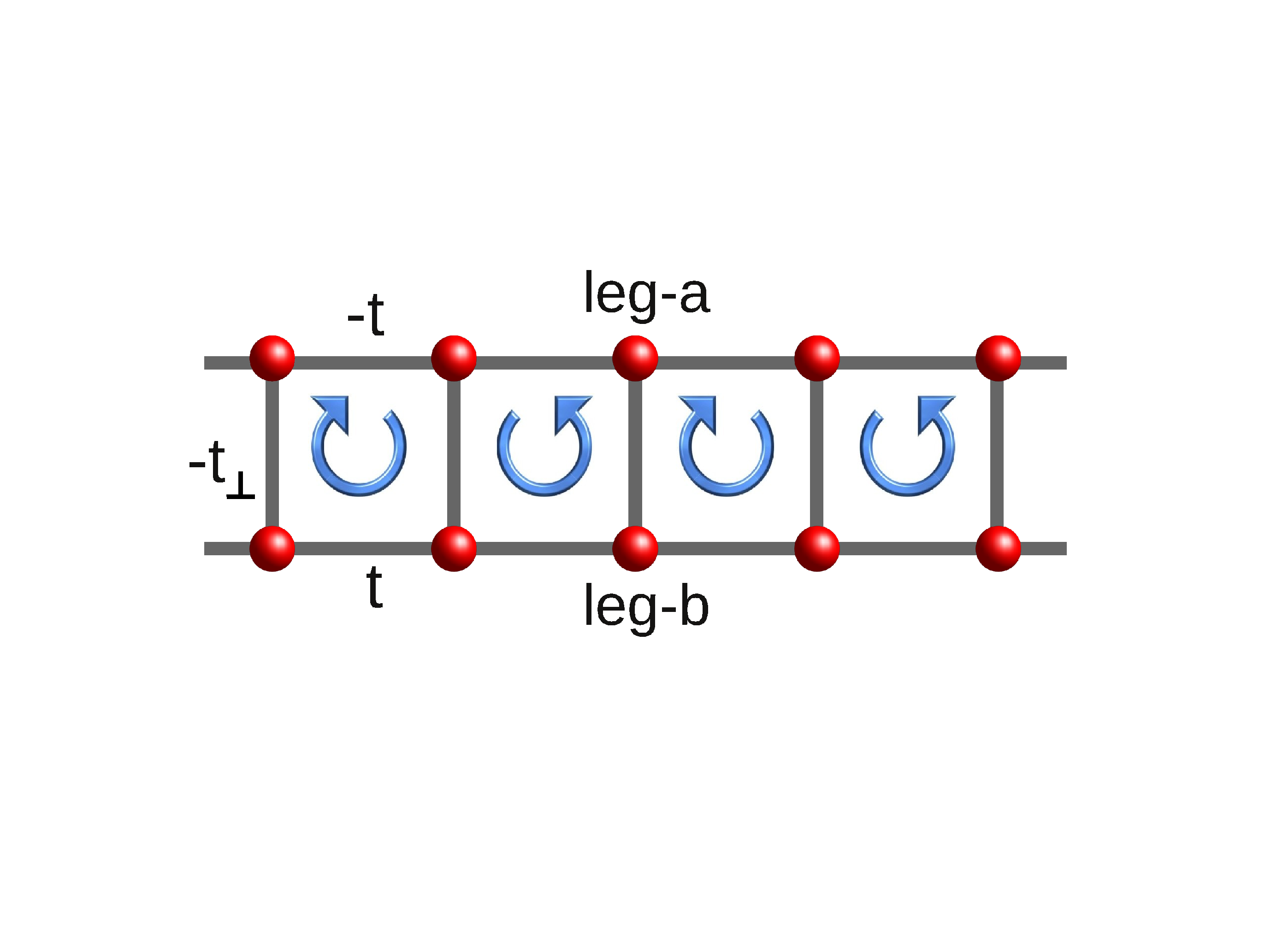}
\vskip -0.4in
\caption{Sketch of the fully frustrated Bose Hubbard two-leg ladder. The opposite signs of the
hopping on the $a$ and $b$ legs correspond to
a $\pi$-flux threading each square plaquette and frustrating the boson kinetic energy. Also
depicted at the spontaneously generated staggered loop currents generated in the superfluid
and chiral Mott insulator phases of this ladder model.}
\label{fig:ladder}
\end{figure}

The Hamiltonian for the frustrated ladder as shown in Fig.~\ref{fig:ladder} can be written as
\begin{eqnarray}
H \!&=&\!\! -t \sum_x (a^\dagger_x a^{\phantom \dagger}_{x+1} \!+\! a^\dagger_{x+1} a^{\phantom \dagger}_{x})
\!+\! t \sum_x (b^\dagger_x b^{\phantom \dagger}_{x+1} \!+\! b^\dagger_{x+1} b^{\phantom \dagger}_{x}) \nonumber \\
\!&-&\!\! t_\perp
\sum_x (a^\dagger_x b^{\phantom \dagger}_{x} + b^\dagger_{x} a^{\phantom \dagger}_{x}) + \frac{U}{2} \sum_x (n^2_{a,x} + n^2_{b,x}),
\label{LadderHam}
\end{eqnarray}
where $a_x$ and $b_x$ are bosonic operators on each of the two legs of the ladder whose sites are labelled by $x$.
$n_{a,x}$ and $n_{b,x}$ are the corresponding occupation numbers. $t_\perp$ is the hopping amplitude between the legs and $U$ is
the onsite two-body interaction strength.

In our previous work we studied the ground state of the Eqn.~\ref{LadderHam} at integer filling
using two independent numerical methods such as
classical Monte Carlo techniques\cite{dhar.pra2012} and density matrix renormalization group(DMRG).
The complete phase diagram of this model is
shown in Fig(\ref{fig:phasedia-mc}) and Fig(\ref{fig:phasedia-dmrg}) and described in detail in the corresponding sections.
The important results obtained by the above two analysis are
qualitatively similar. As a result of the competition between
the onsite interaction ($U$), intrachain hopping( $t$) and
interchain hopping ($t_\perp$), we obtain {\it three} different quantum phases: the CSF, the CMI and a regular MI . When the Hubbard repulsion ($U$) is small the system exhibits a gapless SF phase with a finite loop current order in
each plaquette. For intermediate values of $U$ the system undergoes a transition from CSF to CMI phase which possesses finite charge gap and
also exhibits staggered loop current and spontaneously breaks time reversal symmetry. A further increase of the value of $U$ breaks the loop
current order in the system and the system undergoes a transition to the regular MI phase. Using different scaling properties of the
observables obtained by the numerical simulations we found that the SF to CMI phase transition is of Berezinskii-Kosterlitz-Thouless (BKT) \cite{kosterlitz.jpc1973}
universality class and the transition from CMI to MI is of Ising universality class. Apart from the numerical analysis of this model we analyse the system using
several analytical methods. We explain the numerical and analytical methods used to understand the ground state properties of the Hamiltonian given in Eqn.~\ref{LadderHam}, and describe the properties of the different phases, in the
following sections.

\section{Approximate analyses of the FFBH ladder model}

\subsection{Single particle condensate wavefunction}
\begin{figure}[t]
\includegraphics[width=3.2in]{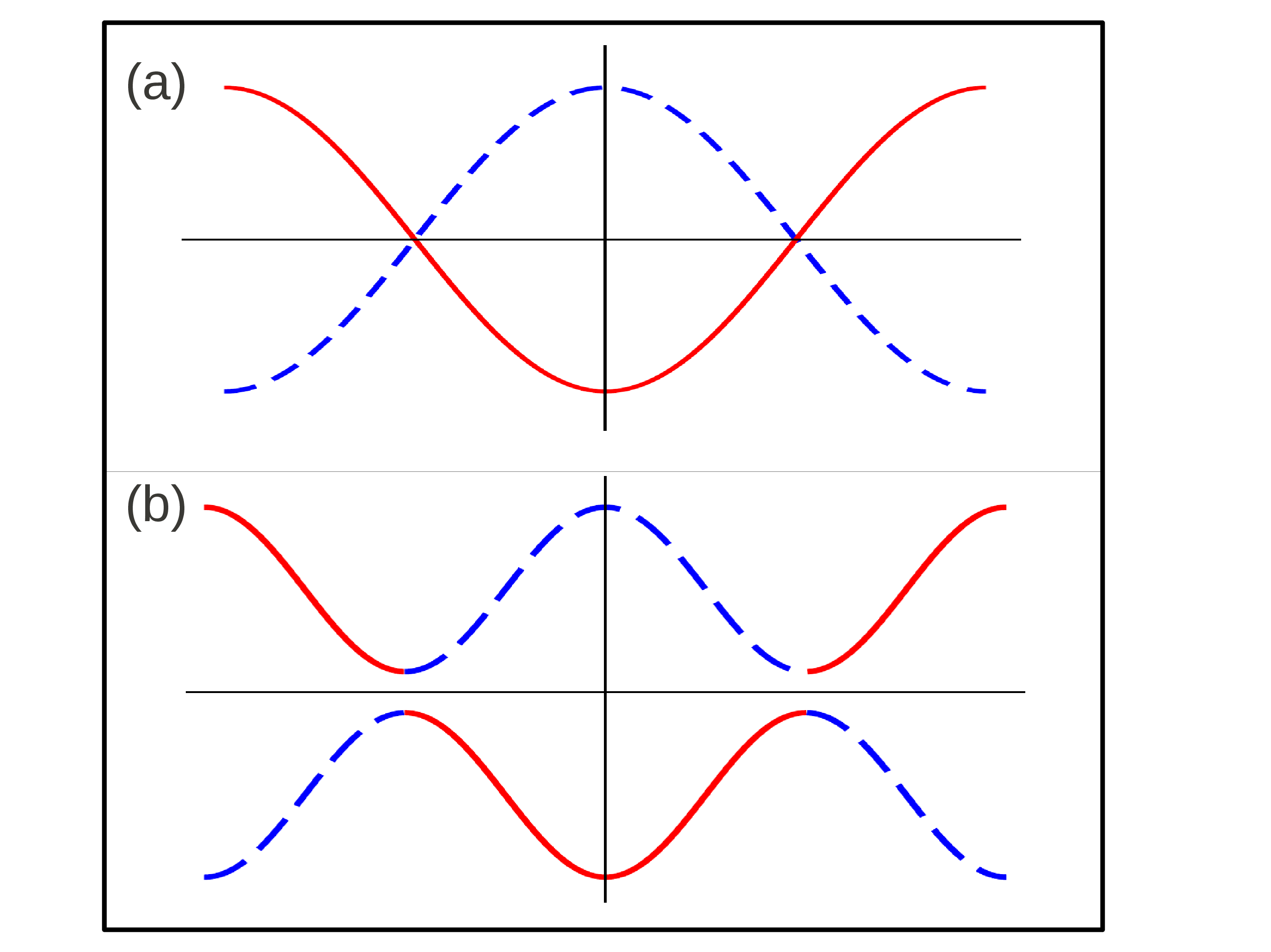}
\caption{(Top) The single particle dispersion obtained from Eqn.~\ref{LadderHam} with $t_\perp=0$. The two legs are now completely decoupled from each other. For $t>0$, the lower (upper) band corresponds to the bosons being in the $a$ ($b$) leg as is shown by the solid (dashed) line.. The two bands are degenerate at $k=\pm \pi/2$. (Bottom) For $t_\perp \neq 0$, the degeneracies between the bands are lifted resulting in a gap between the two bands. There are now two minima in the lower band at $k=0$ and $\pi$. The $k=0$ ($\pi$) minimum originates from the lower (upper) band of the two decoupled chains and thus corresponds to the particles being localized mostly in the $a$ ($b$) legs. }
\label{fig:dispersion1}
\end{figure}

We begin by motivating the form of the single particle condensate wavefunction for our system in the weakly
interacting limit. To do this we first consider the single
 particle dispersion obtained from Eqn.~\ref{LadderHam}, i.e. setting $U=0$. If we also set $t_\perp=0$, the dispersion is
as shown in the upper panel of Fig.~\ref{fig:dispersion1} with two bands inverted with respect to each other and intersecting
at $k=\pm \pi$. The band with a minimum at $k=0$ corresponds to the $a$ leg while the one with a maximum at $k=0$ corresponds
to the $b$ leg. With $t_\perp \neq 0$, the degeneracies at the points of intersection are lifted resulting in two bands with
a gap as shown in the lower panel of Fig.~\ref{fig:dispersion1}. The two minima at $k=0$ and $k=\pi$ originate from the bands
 corresponding to each of the two legs with the $k=0$ minimum corresponding to bosons localized in the $a$ leg and $k=\pi$
 minimum to bosons in the $b$ leg. A single particle condensate wavefunction for $U=0$ can thus be any linear superposition of the states at the two minima.

For $U > 0$ at a filling of one boson per site, the wavefunction has the form
\begin{equation}
|\Psi\rangle = \frac{1}{\sqrt{2}} e^{i\theta}\left(|k=0\rangle + e^{i\phi}|k=\pi \rangle \right),
\label{Eq:conwavefn1}
\end{equation}
where $|k=0\rangle$ and $|k=\pi \rangle$ correspond to the states at the minima $k=0$ and $k=\pi$ respectively and $\theta$ and $\phi$ are phases. This form of $|\Psi\rangle$ comes about because the onsite repulsion discourages double occupancy so that with one boson per site, the system favors distributing the bosons equally over both legs. Thus, the single particle condensate wavefucntion has equal amplitude in the states $|k=0\rangle$ and $|k=\pi \rangle$ with relative phase $\phi$ and global phase $\theta$. Minimizing the energy of states of the form given in Eqn.~\ref{Eq:conwavefn1} yields $\phi = \pm \pi/2$. Thus, the form of the single particle wavefunction is
\begin{equation}
|\Psi\rangle = \frac{1}{\sqrt{2}} e^{i\theta}\left(|k=0\rangle + e^{\pm i\pi/2}|k=\pi \rangle \right),
\label{Eq:conwavefn}
\end{equation}
which has a global $U(1)$ symmetry corresponding to $\theta$, which acts as the superfluid parameter and a $Z_2$ symmetry due to the $\pm$ sign in the relative phase corresponding to a particular pattern of staggered current loops (chirality). In 1 + 1 D, like for our system, there is no long range $U(1)$ order but at best algebraically decaying correlations corresponding to quasi-long range order of the superfluid. However, there can exist true long range Ising order. The state with coexisting algebraic superfluid order and long range chiral order is the chiral superfluid. The chiral Mott insulator corresponds to short range superfluid correlations (accompanied by a charge gap) existing along with long range chiral order whereas in the Mott insulator both superfluid and chiral order are completely absent.

\subsection{Mean field theory of the chiral superfluid}

In this section we refine the above discussion and present the mean-field theory description of the chiral superfluid phase in (\ref{LadderHam}).
Ignoring the interaction term $U$,
we obtain the single particle dispersion $\pm E_k$ with $E_k= \sqrt{t_\perp^2 + (2 t \cos k)^2}$ and
introducing a chemical potential $\mu$, the two bands have dispersion $\pm E_k - \mu$. Of these,
the lower band (with dispersion $-E_k-\mu$) has two minima at $k=0, \pi$. The eigenvector for this band $(u_k,v_k)^T$ is
\bea
u_k &=& \frac{1}{\sqrt{1+ g_k^2}} = \sqrt{\frac{1}{2} (1  + \frac{2 t \cos k}{E_k})} \label{equk} \\
v_k &=& \frac{g_k}{\sqrt{1+ g_k^2}} = \sqrt{\frac{1}{2} (1  - \frac{2 t \cos k}{E_k})} \label{eqvk} \\
u_k v_k &=& \frac{t_\perp}{2 E_k}
\eea
where $g_k = (E_k-2 t)/t_\perp$.
Note that we can set $u_{k+\pi}=v_k$ and $v_{k+\pi} = u_k$ which means, in particular, that $u_\pi=v_0$ and $v_\pi=u_0$.

We can rotate to a new basis, where $\alpha^\dg_k$ ($\beta^\dg_k$) creates quasiparticles in the lower
(higher) band at momentum $k$, via
\be
\begin{pmatrix} a^\pdg_k \\ b^\pdg_k \end{pmatrix} = \begin{pmatrix} u_k & -v_k \\ v_k & u_k \end{pmatrix}
\begin{pmatrix} \alpha^\pdg_k \\ \beta^\pdg_k \end{pmatrix}
\ee in which the single particle Hamiltonian is diagonal and given
by \be H_{\rm kin} = \sum_k \left[ (-E_k - \mu) \alpha^\dg_k
\alpha^\pdg_k + (E_k - \mu) \beta^\dg_k \beta^\pdg_k) \right]. \ee
Reintroducing the local repulsive interaction $H_{\rm int} =
\frac{U}{2} (n^2_a + n^2_b)$ at each site, which is responsible for
eventually driving the 2-leg ladder into a Mott insulator state, we
get back the Hamiltonian in Eqn.~\ref{LadderHam}. We focus on the
low energy physics and thus ignore any effects of the upper band.
Further, at the mean-field level we can focus only on the lowest
energy $k=0,\pi$ modes in the low energy band, and write the
Hamiltonian in terms of these modes. This leads to
\begin{eqnarray}
H^{\rm proj}_{\rm low} &=& (-E_0 - \mu)\sum_{i=0,\pi} \alpha^\dg_i \alpha^\pdg_i \nonumber \\
&&+ U (u_0^4 + v_0^4) \sum_{i=0,\pi} \alpha^\dg_i \alpha^\dg_i \alpha^\pdg_i \alpha^\pdg_i \nonumber \\
&&+8 U u_0^2 v_0^2 \alpha^\dg_0 \alpha^\dg_\pi \alpha^\pdg_\pi \alpha^\pdg_0  \nonumber \\
&&+2 U u_0^2 v_0^2 (\alpha^\dg_0 \alpha^\dg_0 \alpha^\pdg_\pi \alpha^\pdg_\pi +
\alpha^\dg_\pi \alpha^\dg_\pi \alpha^\pdg_0 \alpha^\pdg_0 )
\end{eqnarray}

If we were to condense the $\alpha_i$ bosons with $\la \alpha_i \ra = \varphi_i$ (where $i=0,\pi$), we get
the mean field energy
\bea
E^{\rm mft}_{\rm low} = (-E_0 - \mu) \sum_{i=0,\pi} |\varphi_i|^2
+ U (u_0^4 + v_0^4) \sum_{i=0,\pi} |\varphi_i|^4 \nonumber \\
+ 8 U u_0^2 v_0^2 |\varphi_0|^2 |\varphi_\pi|^2 + 2 U u_0^2 v_0^2
(\varphi_0^{* 2} \varphi_\pi^2  +  \varphi_\pi^{* 2} \varphi_0^2),
\eea
where the final term describes ``Umklapp effects'' which transfers a pair
of bosons from one minimum to the other.
Minimizing the interaction energy with respect to the phase
difference between $\varphi_0$ and $\varphi_\pi$ gives a value of
$\pm \pi/2$ phase for this quantity.
 Minimizing with respect to the amplitude of the two condensates we find it to be the same for
 both, which we call $\psi$. This gives us up to a global phase rotation,
\bea
\la a_x \ra &=& \psi \left[ u_0  \pm i v_0 (-1)^x \right]\\
\la b_x \ra &=& \psi \left[ v_0 \pm  i u_0 (-1)^x \right]. \eea This
leads to a spatially uniform boson density and bond energy, but to a
spatially varying bond current density which forms a staggered
pattern, much like a vortex-antivortex crystal. Specifically, the
bond currents are given by \bea
j^a_{x,x+1} &=& -  i t \la a^\dg_x a^\pdg_{x+1} - a^\dg_x a^\pdg_{x+1} \ra \nonumber \\
&=& \mp 4 t \psi^2 u_0 v_0 (-1)^x \\
j^b_{x,x+1} &=& +  i t \la b^\dg_x b^\pdg_{x+1} - b^\dg_x b^\pdg_{x+1} \ra \nonumber \\
&=& \pm 4 t \psi^2 u_0 v_0 (-1)^x \\
j^{ab}_{x} &=& -  i t_\perp \la a^\dg_x b^\pdg_{x} - b^\dg_x a^\pdg_{x} \ra \nonumber \\
&=& \mp 2 t_\perp \psi^2 (u_0^2 - v_0^2) (-1)^x. \eea Note that
current conservation at each lattice point works out if we use the
fact, coming from the single particle dispersion analysis, that $4 t
u_0 v_0 = t_\perp (u_0^2 - v_0^2)$. The two possible choices for the
sign of each bond current correspond to two distinct current order
patterns which are related to one another by time-reversal or by a
unit translation.

\subsection{Mean field theory of the Mott transition}

We next turn to the Mott insulating state induced by strong repulsion in the CSF state. We
describe here the mean field theory of this Mott transition, which we then argue to be
inadequate to describe the physics. To obtain the mean field CSF to insulator phase
boundary, we consider the single site
Hamiltonian (motivated by our previous discussion of the CSF),
\bea
H_{\rm mf}^{\rm single}(a) &=& - 2 t \psi ((u_0 - i v_0) a^\dg + (u_0 + i v_0) a^\pdg) \nonumber \\
&-& t_\perp \psi ((v_0 + i u_0) a^\dg
+ (v_0 - i u_0) a^\pdg) \nonumber \\
&+& \frac{U}{2} a^\dg a^\dg a^\pdg a^\pdg - \mu a^\dg a^\pdg. \eea
Setting $t$ and $t_\perp$ to 0 we obtain a single site Hamiltonian,
whose ground state is a true number state $| n \ra$ given by $ U (n
- 1) < \mu < U n$. Turning on a small nonzero $\psi$ leads to the
first order corrected wavefunction \be |\tilde n^{(1)} \ra = | n \ra
+ \frac{\psi (r - i s) \sqrt{n+1}}{U n - \mu} |n+1\ra + \frac{\psi
(r + i s) \sqrt{n}}{\mu - U (n-1)} | n -1 \ra, \ee where \bea
r &\equiv& (2 t u_0 + t_\perp v_0), \\
s &\equiv& (2 t v_0 - t_\perp u_0).
\eea
The resulting linearized self-consistent equation for $\psi$ leads to the phase boundary
\be
\frac{1}{\sqrt{4 t^2 + t^2_\perp}} = \frac{n}{\mu-U(n-1)} + \frac{n+1}{U n - \mu}
\ee
with $U(n-1) < \mu <  U n$.

This result for the SF-MI phase boundary
is exactly the same as in the usual mean field SF-MI phase diagram provided we replace
$z t \to \sqrt{4 t^2 + t^2_\perp}$ in the conventional result (where $z$ is the coordination number),
so that the Mott transition happens when
\be
\frac{U_{c,ladder}^{\rm \pi-flux}}{\sqrt{4 t^2 + t^2_\perp}} = g_\ast
\label{Eq:mftrans}
\ee
where $g_\ast \approx 5.83$ at a filling of one boson per site.

We end by noting that this mean field theory does not
have explicit current density wave and superfluidity as two independent order parameters, since the current is
simply a product of the Bose condensate at two different sites. Thus the superfluidity and any time-reversal
breaking orders vanish together at a single SF-MI transition. However, such a vanishing of two completely
different orders at a single continuous Mott transition is not expected to be generic and is Landau-forbidden.
This encourages us to pursue careful numerical studies of this model which enable us to go beyond this simple
mean field theory.

\section{classical XY model}

As noted above, mean field theory fails to give us the complete picture of the FFBH model. In order to go beyond
this approach, we map the quantum
FFXY model (which is a good description of the FFBH model at large integer filling) to an effective classical model
on a space-time lattice by stacking of ladders atop each other in the imaginary space time. To do this we construct a
rotor model from the fully frustrated Bose Hubbard model in Eqn.~\ref{LadderHam} -
such a rotor model ignores amplitude
fluctuations and is expected to be a valid effective description at large fillings. Setting $a^\pdg_x \sim \exp(-i \varphi_x^a)$ and
$b^\pdg_x \sim \exp(-i \varphi_x^b)$ and replacing the number operators $n^{a,b}$ by angular momentum operators
${\cal L}^{a,b}$ which cause fluctuations in the phases (angles) $\varphi^{a,b}$ we obtain the partition function in terms of the
classical action in one higher dimension as
\be
Z = \sum_{\{\{\varphi_{x,\tau}\}\}} {\rm e}^{-S^{1+1}_{\rm cl}[\varphi]}
\ee
where
\begin{widetext}
\bea
S^{1+1}_{\rm cl} &=&  - \sum_{x \tau} \left[J_\parallel \cos(\varphi^a_{x+1,\tau} - \varphi^a_{x,\tau}) -
J_\parallel \cos(\varphi^b_{x+1,\tau} -
\varphi^b_{x,\tau}) + J_\perp \cos(\varphi^a_{x,\tau} - \varphi^b_{x,\tau}) \right] \nonumber \\
&-& J_\tau \sum_{x \tau}
\left[ \cos(\varphi^a_{x,\tau+1} - \varphi^a_{x,\tau}) + \cos(\varphi^b_{x,\tau+1} - \varphi^b_{x,\tau}) \right],
\eea
\end{widetext}
with $2 \epsilon \tt=J_\parallel$, $2 \epsilon \tt_\perp = J_\perp$, and $1/\epsilon U = J_\tau$ (see Appendix A for 
a detailed derivation).
We see that this has the form of a $XY$ model.
\begin{equation}
H_{\rm XY} = - \sum_{i, \delta} J_\delta \cos \left( \varphi_i - \varphi_{i+\delta}\right).
\label{Eq:classham}
\end{equation}
Here the classical variable $\varphi_i$ corresponds to the boson
phases and $(i,i\! +\! \delta)$ are the nearest neighbours along the
space-time direction $\delta$. The couplings $J_\delta$ take the
values $\pm J_\parallel$ on the two legs, $J_\perp$ on the rungs
linking the two layers, and $J_\tau$ in the imaginary time
direction. In order to get the properties of the quantum model at a
fixed inverse temperature $\beta \tt$, we must take the
`time'-continuum limit of $\epsilon \to 0$, sending $J_\parallel \to
0$, $J_\perp \to 0$, and $J_\tau \to \infty$, while keeping fixed
$J_\perp/J_\parallel = \tt_\perp/\tt$  and $J_\tau J_\parallel = 2
\tt/U$. The inverse temperature $\beta \tt$ is then given by
$\epsilon \tt L_\tau$ and thus depends on the chosen value of
$\epsilon \tt$ (which must be taken to be very small) and the size
of the simulation cell in the `time'-direction. We set $\epsilon =
1/\sqrt{2 U \tilde{t}}$ which leads to $J_\parallel = J_\tau =
\sqrt{2 \tilde{t}/U}$.

\begin{figure}[t]
\includegraphics[width=3.2in]{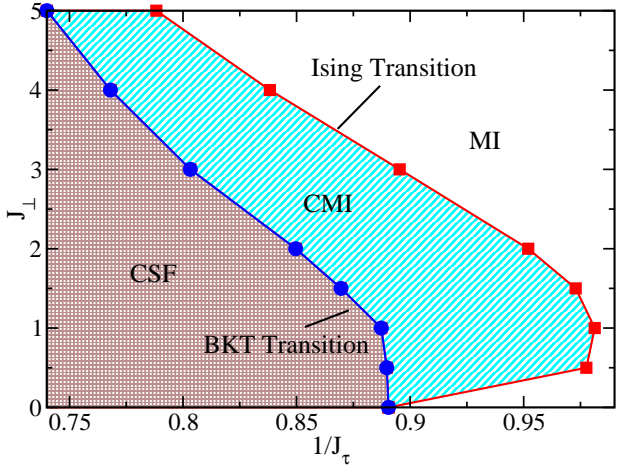}
\caption{Phase diagram of the effective classical model obtained in ref.~\onlinecite{dhar.pra2012}
from classical Monte Carlo simulations of Eqn.~\ref{Eq:classham}, which is shown in with
$J_\parallel = J_\tau$, obtained from variational Monte Carlo simulations. }
\label{fig:phasedia-mc}
\end{figure}

This effective square lattice bilayer XY Hamiltonian for the FFBH model
can be simulated using a classical Monte Carlo algorithm.  The Monte Carlo simulation was performed using the Metropolis algorithm
with selective energy conserving moves to increase the acceptance rates at large coupling. About $10^6 - 10^7$ steps were used to calculate
the equilibrium values of the various quantities at the largest system sizes.

The ordinary Mott insulator of
the FFBH model corresponds to a fully disordered paramagnetic state of the
effective bilayer classical XY model. Starting from the
superfluid phase at small $1/J_\tau$, we detect the vanishing of superfluidity in this
model by
calculating the helicity modulus $\Gamma$, which corresponds to a response to
an infinitesimal phase twist in the spatial direction. Explicitly,
\be
\Gamma = \left. \frac{1}{2} \frac{\partial^2 F}{\partial \Phi^2} \right|_{\Phi \rightarrow 0},
\ee
where the free energy
\be
F = -\log \sum_{\{\varphi_{x,\tau}\}} e^{-S^{1+1}_{\rm cl}},
\ee
and $\Phi$ is the flux twist along the $\parallel$ direction.

\begin{figure}[t]
\includegraphics[width=3.4in]{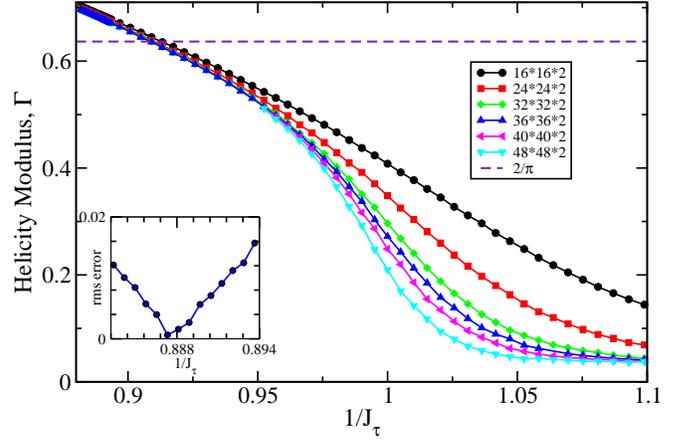}
\caption{ Helicity modulus $\Gamma$ versus $1/J_\tau$ for different system sizes
for $J_\perp\!=\!1$ as also obtained in ref.~\onlinecite{dhar.pra2012}. (Inset) RMS error of fit to the BKT finite
size scaling form of $\Gamma$
shows a deep minimum at the transition, at $1/J_\tau \!=\! 0.887(1)$, and yields a jump
$\Delta \Gamma\! \approx\! 0.637$,
close to the BKT
value of $2/\pi$. }
\label{Fig:helicityplot}
\end{figure}

As seen from Fig. \ref{Fig:helicityplot}, $\Gamma$ appears to drop
precipitously in the range $1/J_\tau \sim 0.8-1.1$, with the size
dependence of $\Gamma$ indicative of a jump in the thermodynamic
limit, as expected at a BKT transition. However, for finite-sized
systems, this jump is rounded off and one has to use finite-size
scaling to precisely locate the transition. The relevant equation
which governs the behavior of the helicity modulus at the transition
point is obtained from integrating the Kosterlitz-Thouless
renormalization group equations, and
yields~\cite{weber_minnhagen,olsson.prl1995} \be \Gamma (L) = A
\left( 1 + \frac{1}{2} \frac{1}{\log L + C}\right),
\label{Eq:BKTfinite} \ee where the constant $A$ has the universal
value of $2/\pi$ for the BKT transition while $C$ is a non-universal
constant and the classical system is of size $L \times L \times 2$
(with the factor of $2$ for our bilayer). $\Gamma (L)$ is calculated
numerically for different values of the parameter being varied to
effect the transition and fit to Eqn.~\ref{Eq:BKTfinite}. The value
of the parameter which gives the best fit is where the transition
takes place. In Fig.(\ref{Fig:helicityplot}) we show that the BKT
transition point from CSF to CMI is at $1/J_\tau \!=\! 0.887(1)$ for
$J_\perp\!=\!1$. This method has been used to successfully locate
the thermal BKT transition for a single layer XY model,
with~\cite{olsson.prl1995} or without
frustration~\cite{weber_minnhagen}, and has also been used to detect
non-BKT thermal transitions driven by half-vortices (which yield
$A=8/\pi$) in spinor condensates~\cite{mukerjee_prl}. Using this
technique for several values of $J_\perp$ we obtain the boundary for
the CSF-CMI transition as shown in Fig. \ref{fig:phasedia-mc}.

\begin{figure}[!t]
\includegraphics[width=3.4in]{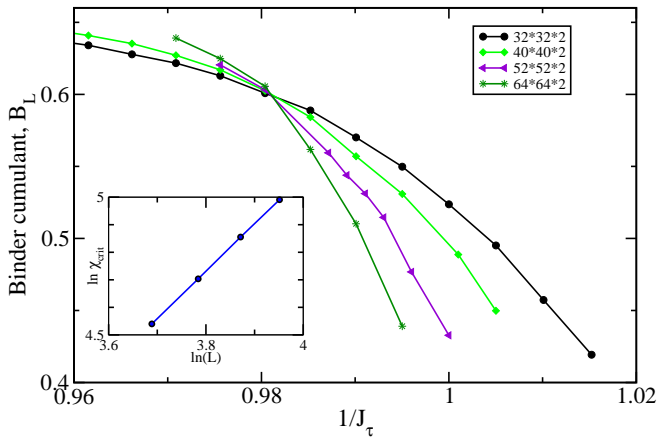}
\caption{Binder cumulants for the staggered current versus $1/J_\tau$ (for different $L$
for $J_\perp\!=\!1$) intersecting at a continuous transition at $1/J_\tau\! =\! 0.981(4)$ as also obtained in
ref.~\onlinecite{dhar.pra2012}.
(Inset) Critical susceptibility versus $L$ gives the ratio of critical exponents $\gamma/\nu \! \approx\! 1.72$,
very close to 2D Ising value $\gamma/\nu\!=\!7/4$. }
\label{Fig:binderplot}
\end{figure}

The Ising transition requires a different method for its detection.
Since the chiral order is truly long ranged, we can use the method
of Binder cumulants to accurately locate the transition. For an
order parameter $m$, the Binder cumulant is defined as \be B_L =
\left(1 - \frac{\langle m^4 \rangle_L}{3 \langle m^2 \rangle^2_L}
\right). \label{Eq:binder} \ee For our system, the order parameter
is given by \be m = \frac{1}{L^2}\sum_{i\tau} \left( -1 \right)^i
J_{i\tau}, \label{Eq:IsingOP} \ee where $J_{i\tau}$ is the current
around a plaquette normal to the time direction. $i$ and $\tau$ are
respectively the coordinates in the $\parallel$ and time directions.
If $a(i,\tau)$, $a(i+1, \tau)$, $b(i+1, \tau)$ and $b(i, \tau)$ are
the vertices of the plaquette going around clockwise, \bea
J_{i\tau}& =& J_\parallel \left[ \sin \left(\varphi^a_{i+1,\tau} - \varphi^a_{i, \tau} \right) + \sin \left(\varphi^b_{i,\tau} - \varphi^b_{i+1, \tau} \right) \right] \nonumber \\
&+& J_\perp \left[ \sin \left(\varphi^b_{i+1,\tau} - \varphi^a_{i+1,
\tau} \right) + \sin \left(\varphi^a_{i,\tau} - \varphi^b_{i, \tau}
\right) \right]. \label{Eq:plaqcurrent} \eea The order parameter
given by Eqns.~\ref{Eq:IsingOP} and \ref{Eq:plaqcurrent} is like a
staggered magnetization generated by the loop currents. Since the
current in a given loop can be clockwise or counterclockwise, this
order parameter is of the Ising type. The values of $B_L$ calculated
for different $L$ are equal only at the fixed points of the system.
In addition to the low and high temperature fixed point, they will
be equal at the location of a continuous transition separating an
ordered phase from a disordered phase. Thus, plotting $B_L$ as a
function of the tuning parameters for different values of $L$
reveals the location of the transition at the point where the curves
intersect as shown in Fig. \ref{Fig:binderplot}, where we plot $B_L$
as a function of $1/J_\tau$ for $J_\perp\!=\!1$. Curves for
different $L$ intersect at $1/J_\tau\! =\! 0.981(4)$ which is the
transition point from CMI to MI phase.

It is important to note that the method of Binder cumulants does not assume any particular universality class for the transition.
To verify that the transition in question is indeed of the Ising universality class,
as expected on the grounds that the CMI breaks a two-fold discrete translational symmetry (or
time reversal symmetry,
we have calculated the susceptibility $\chi$ for the onset of chiral order. At the transition,
$\chi$ is expected to diverge with system size as $\chi \sim L^{\gamma/\nu}$, where $\gamma$ and $\nu$ are respectively,
the conventional susceptibility and correlation length critical exponents. For the classical 2D Ising universality class,
$\gamma/\nu=7/4$, while we find $\gamma/\nu \approx 1.72$(inset of Fig.\ref{Fig:binderplot}),
 which demonstrates that our transition is indeed in the Ising universality class. By performing this proceedure for several values of $J_\perp\!=\!1$ we
obtain the Ising transition line as shown in the Fig. ~\ref{fig:phasedia-mc}.

In the previous studies of $H_{\rm XY}$ it has been shown that there
exists only one transition when $J_\perp=J_\parallel$ and for large
anisotropy the system exhibits two separate
transitions.\cite{granato.prb1993,Granato2} In our study, we find
the CMI phase at the fully symmetric point
$J_\perp=J_\parallel=J_\tau$ indicating two transitions. The absence
of the second phase transition in the previous study could be due to
the small system sizes considered.

\section{Density matrix renormalization group study}
We have performed a finite size density matrix renormalization group (FS-DMRG)
study on the model described by Eqn.~\ref{LadderHam} to understand different quantum
phase transitions and compute an accurate phase diagram.\cite{whitedmrg.prl1992,schollwock.rmp2005}
In our FS-DMRG calculation we have mapped the 2-leg Bose-ladder into
a single chain with appropriate hopping elements. Calculations were performed
up to a system length of $200$ sites (which is equivalent to $100$
rungs of the ladder system) at a filling one boson per site.
We keep six states per site and retain $200$ density matrix states in our calculation.
The error due to the weight of the discarded states is expected to be less than $10^{-5}$. By calculating different
relevant physical quantities we obtain an accurate ground state phase diagram of Eqn.~\ref{LadderHam} as explained below.

\begin{figure}[t]
\includegraphics[width=3.2in]{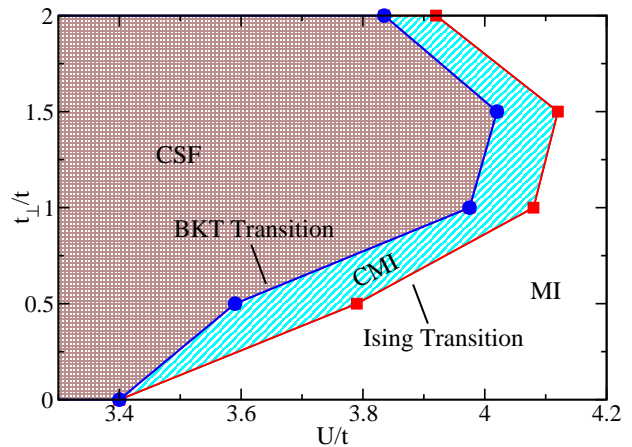}
\caption{ Phase diagram of the FFBH model in Eqn.~\ref{LadderHam} obtained using DMRG in
ref. ~\onlinecite{dhar.pra2012}.
Both the classical and quantum models exhibit
a chiral Mott insulator (CMI) at intermediate correlations, intervening between a chiral superfluid (CSF)
and an ordinary Mott insulator (MI).}
\label{fig:phasedia-dmrg}
\end{figure}

We analyze the CSF-CMI and CMI-MI transitions
using a finite-size analysis of the momentum distribution function and the rung current
structure factor respectively. These are defined as
\begin{equation}
n(k)=\frac{1}{L}\sum_{x,x'}e^{ik(x-x')} \left[ \langle a_x^{\dagger} a_{x'} \rangle + \langle b_x^{\dagger} b_{x'} \rangle \right]
\label{eq:mom}
\end{equation}
and
\begin{equation}
S_j(k)=\frac{1}{L^2}\sum_{x,x'}{e^{ik(x-x')}\langle{j_x j_{x'}}\rangle},
\label{eq:current}
\end{equation}
where $j_x=i\left( a_x^\dagger b_x - b_x^\dagger a_x \right)$.

\begin{figure}[!t]
\begin{center}
\includegraphics[height=2in]{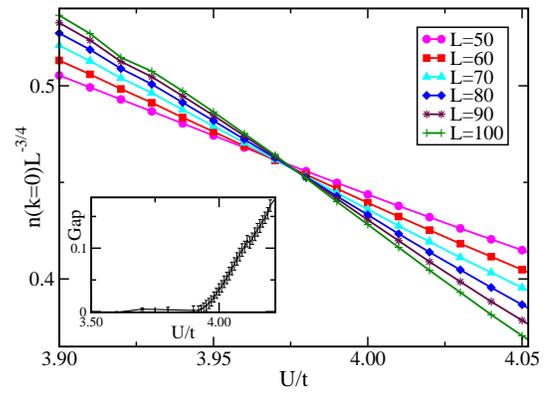}
\end{center}

\vspace*{0.25cm}
\begin{center}
\includegraphics[height=2in,width=0.47\textwidth]{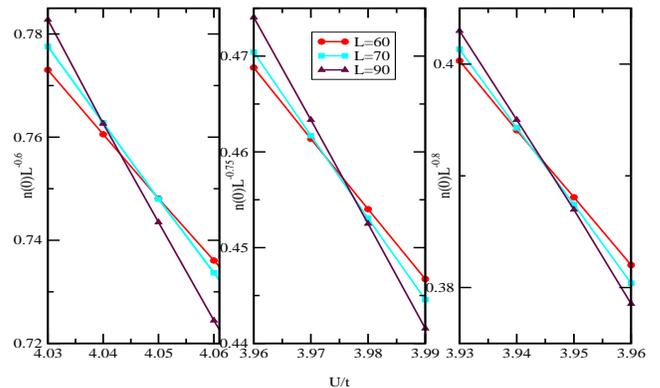}
\caption{Top panel: $n(0)L^{-3/4}$ as a function of $U/t$ for different
system sizes, showing a crossing point at $U_{c,1}/t=3.98(2)$
which we identify as the CSF-CMI
transition point which is in the BKT universality class. The inset shows
the charge gap opening up at this transition into the insulator.
Bottom panel: $n(0)L^{-\alpha}$ as a function of $U/t$ for three
different values of $\alpha = $ 0.6 (left), 0.75 (middle) and 0.8
(right) and different system sizes. The curves can be seen to cross
sharply for $\alpha = 0.75$ and 0.8 whereas they do not for
$\alpha=0.6$.}
\end{center}
\label{fig:n0_plots}
\end{figure}

At a BKT transition,
\be
\langle a_x^\dagger a_{x'}\rangle = \langle b_x^\dagger b_{x'}\rangle \sim 1/|x-x'|^{1/4},
\ee
 so $n(0)L^{-\alpha}$ with $\alpha=3/4$ is independent of the system length $L$.
Thus curves of $n(0)L^{-3/4}$ as functions of the tuning parameter for different $L$ will intersect at a point which we have used to locate the transition as
shown in Fig.~\ref{fig:n0_plots}. In the top panel of Fig.~\ref{fig:n0_plots} we plot $n(k=0)L^{-3/4}$
versus $U/t$  with $t_\perp\! =\! t$. The curves for different $L$
intersect at $U_{c1}/t \approx 3.98(1)$ showing the CSF-CMI transition. The inset shows that the charge gap
in the insulator also becomes nonzero at this point, thus providing a nontrivial consistency check that we
have correctly located the superfluid to insulator transition.

In order to see that such a
intersection is not
obtained for $\alpha < 3/4$ (corresponding to real space correlations decaying with a power law
faster than $1/|x-x'|^{1/4}$), we plot, in the bottom left panel of Fig.~\ref{fig:n0_plots}, $n(0) L^{-\alpha}$ with
$\alpha=0.6$ and find that the
crossing point of these curves for different system sizes drifts significantly instead of coinciding.
For example, the crossing point
between $L=70$ and $L=90$ is at $U/t=4.040$, where as it shifts to $U/t=4.048$ between $L=60$ and $L=70$.
Within the CSF state, however, we expect real space correlations to decay slower than $1/|x-x'|^{1/4}$,
so that curves of $n(0) L^{-\alpha}$ with $\alpha > 3/4$ are simply expected to cross at smaller
$U$ (i.e., deeper into the CSF and away from the CSF-CMI transition). This is also shown in the bottom panels
of Fig.~\ref{fig:n0_plots}, where the crossing point is much better defined for $\alpha=0.75$ (bottom middle panel, critical
point) and $\alpha=0.8$ (bottom right panel, CSF state).

To summarize, the study of the momentum distribution shows that the ground state
of the FFBH model has critical superfluid correlations, with $\alpha > 3/4$, but that there is
no such critical superfluid for $\alpha < 3/4$, suggesting that $\alpha=3/4$ is the endpoint
of a line of critical points. Finally, the CSF-CMI transition located in this manner from the momentum distribution
completely agrees with the superfluid to insulator transition inferred from the onset of a charge
gap shown in the inset of Fig. ~\ref{fig:n0_plots}.
Such a careful analysis thus unambiguously shows that CSF-CMI transition is of the
usual BKT type.

\begin{figure}[!t]
  \centering
\includegraphics[width=0.47\textwidth]{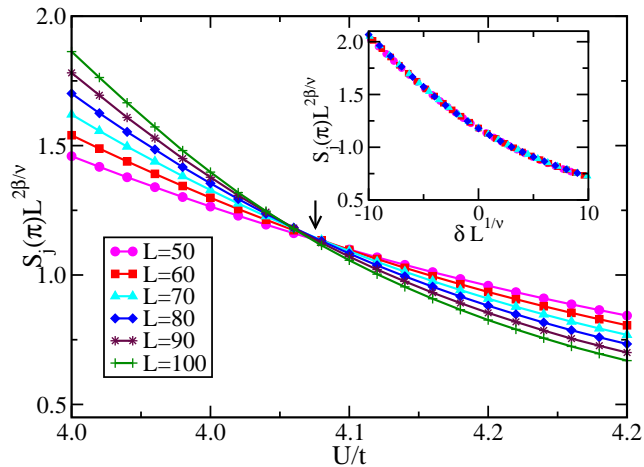}
\caption{ Rung current structure factor $S_j(\pi)L^{2\beta/\nu}$ versus $U/t$ at $t_\perp=1$.
The intersection point yields the CMI-MI Ising
transition at $U_{c2} \approx 4.08(1) t$ as also obtained in ref.~\onlinecite{dhar.pra2012}.
Inset shows $S_j(\pi)L^{2\beta/\nu}$ versus $\delta L^{1/\nu}$ with $\delta \equiv
(U - U_{c 2})/t$, plotted for different $U/t$, leading to a scaling collapse for 2D Ising
exponents $\nu=1$ and
$\beta=1/8$.}
\label{fig:spicrossing}
\end{figure}

Having located the CSF-CMI transition, we next turn to the CMI-MI transition. Based on
the classical model study, the CMI-MI transition is expected to be an Ising transition.
In order to locate this transition for the FFBH model using DMRG, we use
the scaling form of the rung current structure factor $S_j(\pi)$. Since the chiral order is a
staggered pattern of current loops, the ordering vector is at $k=\pi$. To obtain the critical point we use the full scaling form
\begin{equation}
S_j(\pi)L^{2\beta/\nu}=F\left(\left(U-U_{c 2} \right)L^{1/\nu}\right)
\end{equation}
where $\beta$ and $\nu$ are respectively the order parameter and correlation length critical exponents,
and $U$ is the tuning parameter with the transition located at $U_{c 2}$.  For the classical 2D Ising universality class, $\beta = 1/8$ and $\nu =1$.
As a result of the above scaling, curves of $S_j(\pi)L^{2\beta/\nu}$ for different system sizes intersect at the transition point $U_{c2} \approx 4.08(1) t$ as
shown in Fig.~\ref{fig:spicrossing}. Further, with this choice of exponents to rescale the structure
factor and distance to the transition, we obtain a complete data collapse as shown in the
inset of Fig.~\ref{fig:spicrossing}. These are unambiguous
indicators that we have correctly located the CMI-MI transition. The phase diagram obtained using this technique is shown in Fig.~\ref{fig:phasedia-dmrg}.

Finally, in order to show that the CMI is a charge gapped state of bosonic matter, we have computed
the charge gap for $t=t_\perp=1$. The thermodynamic extrapolation of this charge gap is shown in the inset of
the top panel in Fig. \ref{fig:n0_plots}.
The error on the gap is $\approx 0.01$. Again, this unambiguously points to a gapless
CSF state for $U < U_{c1}$, while the states identified as CMI and MI have a nonzero charge gap.

\begin{figure}[t]
  \includegraphics[width=3.2in]{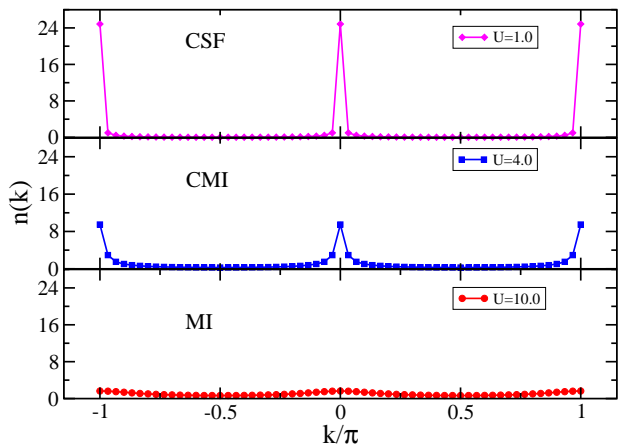}
\caption{Momentum distribution $n(k)$, defined in Eq.~\ref{eq:mom}, computed using DMRG, plotted
in the CSF, CMI, and MI regimes of the phase diagram showing symmetric peaks at $k=0$ and $k=\pi$. While
the CSF exhibits sharp peaks which grow with system size, the CMI and MI phases have finite peaks which
do not scale with system size.}
\label{Fig:nk}
\end{figure}

Using DMRG, we have mapped out the momentum distribution in the various phases - CSF, CMI, and MI.
An analytical discussion of this momentum distribution in the CSF state is given in Appendix B.
In all the phases, we find two peaks at $k=0,\pi$, as seen from the DMRG results shown in Fig.\ref{Fig:nk}. 
While the two peaks in the CSF phase are sharp and grow rapidly with system size,
a careful scaling analysis shows that the peaks in the momentum distribution
do not grow with system size in the CMI and MI.
While this is hard to detect by eye in the
CMI, which shows a sharp peak even for $L=100$, the two peaks are clearly extremely broad in the MI.

\section{Physical pictures for the CMI}

Having identified the remarkable intermediate CMI state using careful numerical studies, we next turn to simple
physical pictures for this state. We first describe the CMI as an exciton condensate approaching it from the Mott
insulator. Next, we argue, from the superfluid side, that the CMI may be alternatively viewed as a vortex supersolid.

\subsection{Exciton Condensate}
Consider the regular Mott insulator state with an integer $n_0 \geq 1$ number of bosons at each site at strong
repulsion $U \gg t, t_\perp$. The low energy excitations about this state correspond to `doublons', obtained by
having an extra particle at a particular site leading to an occupancy $n_0+1$, and `holes', obtained by
removing a particle leading to an occupancy $n_0-1$. The energy cost of a creating a doublon is $U n_0-\mu$,
while the energy cost to create a hole is $\mu - U (n_0-1)$. Once we have a single doublon or hole, this
can move around by boson hopping - the relevant hopping matrix element for these particles can be shown
to be just the bare boson hopping amplitude ($\pm t$, or $t_\perp$) enhanced by a factor of $(n_0+1)$
for doublons and $n_0$ for holes. The resulting low energy bands of doublons and holes thus have energies
\bea
\varepsilon_d(k) &=& U n_0-\mu - (n_0+1) E_k, \\
\varepsilon_h(k) &=& \mu - U (n_0-1) - n_0 E_k,
\eea
where $-E_k$ is the lowest band dispersion of a free single particle on the ladder, given by $-E_k \equiv
- \sqrt{(2 t \cos k)^2 + t^2_\perp}$, which has degenerate minima at $k=0,\pi$.
We can pick $\mu$ such that we have particle-hole symmetry in the sense
that the energy required to create the lowest energy hole (at $k=0,\pi$) is the same as that required to create
the lowest energy doublon (at $k=0,\pi$). This leads to $\mu = U (n_0 - 1/2) - E_0/2$, and
\be
\varepsilon_d(0/\pi) = \varepsilon_h(0/\pi) =  \frac{U}{2} - (n_0 + \frac{1}{2} ) E_0.
\ee
At large $U \gg t, t_\perp$, this energy is positive, so that doublons and holes are suppressed in the Mott
insulating ground state.
A crude estimate of the transition from the Mott insulator to a superfluid phase is given by demanding that
$\varepsilon_{d,h}(0/\pi) = 0$, which yields $U_c(n_0) = (2 n_0 + 1) \sqrt{4 t^2 + t^2_\perp}$, which for
$t_\perp/t=1$ and $n_0=1$ yields $U_c = 6.7 t$, within a factor-of-two of the numerical DMRG result. More
importantly, this calculation shows that the excitations of the regular Mott insulator are similar to gapped
particles and holes in a semiconductor, and the Mott insulator to superfluid transition is similar to metallizing
a semiconductor by reducing its gap. This suggests the possibility of forming doublon-hole bound states, akin
to excitons, in the vicinity of the insulator to superfluid transition where the gap is small. Since our ``Mott
semiconductor'' has multiple minima in its band dispersion, both `direct excitons' and `indirect excitons' are
possible depending on whether the doublon and the hole are at the same momentum (both at $k=0$ or both at $k=\pi$),
or are at different momenta (one at $k=0$ and the other at $k=\pi$).

Insight into the nature of the CMI is obtained by noting that the rung current operator
$j_{ab}(x) = -i t_\perp (a^\dg_x b^\pdg_x - b^\dg_x a^\pdg_x)$ can be re-expressed in the Mott regime as
\begin{widetext}
\be
j_{ab} (x) |{\rm Mott}\rangle
\!\!=\!\! -i t_\perp \sqrt{n_0 (n_0\!+\!1)}  (d^\dg_{a}(x) h^\dg_{b}(x) \!-\! d^\dg_{b}(x) h^\dg_{a}(x)) |{\rm Mott} \ra
\ee
\end{widetext}
where the operators $d^\dg_{a/b}(x)$ and $h^\dg_{a/b}(x)$ create doublons or holes on leg $a/b$
on the rung labelled by $x$, and $|{\rm Mott} \rangle$ denotes a caricature of a Mott state with precisely
$n_0$ bosons at each site. Focusing on the low energy doublon and hole modes amounts to projecting
these creation operators to the lowest dispersing band, and to the vicinity of $k=0,\pi$. This yields
\bea
h^\dg_a(x) &=& u_0 \tilde{h}^\dg_0(x) + v_0 (-1)^x \tilde{h}^\dg_\pi(x) \\
h^\dg_b(x) &=& v_0 \tilde{h}^\dg_0(x) + u_0 (-1)^x \tilde{h}^\dg_\pi(x) \\
d^\dg_a(x) &=& u_0 \tilde{d}^\dg_0(x) + v_0 (-1)^x \tilde{d}^\dg_\pi(x) \\
d^\dg_b(x) &=& v_0 \tilde{d}^\dg_0(x) + u_0 (-1)^x \tilde{d}^\dg_\pi (x)
\eea
where $\tilde{h}^\dg, \tilde{d}^\dg$ are the low energy creation operators in the vicinity of
the indicated momenta. In terms of these, the
rung current operator acting on the Mott state becomes
\be
 j_{ab} (x) |{\rm Mott} \ra =  - i t_\perp (u^2_0 - v^2_0) (-1)^x (\tilde{d}^\dg_0 \tilde{h}^\dg_\pi -
 \tilde{d}^\dg_\pi \tilde{h}^\dg_0) |{\rm Mott} \ra.
\ee
This shows that the current operator behaves as a composite operator that creates an
`indirect exciton'. It suggests that the CMI state, in which the current operator on the rungs
has long range staggered order, may be viewed as an `indirect exciton' condensate,\cite{halperin.rmp1968}
obtained by condensing this composite operator, to yield, $ \la \tilde{d}^\dg_0 \tilde{h}^\dg_\pi -
 \tilde{d}^\dg_\pi \tilde{h}^\dg_0 \ra \sim i \Psi$, in the absence of any doublon/hole condensation, i.e.,
 with $\la \tilde{d}^\dg_{0/\pi} \ra =
\la  \tilde{h}^\dg_{0/\pi} \ra =0$. We note that
the preceding physical insight into the CMI phase does not yield a simple energetic reason for why there should
be such an intermediate phase between the regular Mott insulator and the chiral superfluid. An alternative
scenario, within Landau theory, is that the combined presence of soft
hole/doublon modes and such low energy excitons might
render the transition first order with no intervening CMI state.

\subsection{Vortex Supersolid}
The CSF which possesses staggered loop currents can be understood as a vortex-antivortex crystal. In such a situation, the
vortices and antivortices are generated due to the effect of frustration, and due to the repulsion between them, they arrange themselves in
and antiferromagnetic order. When $U$ is large, the vortices delocalize to form a vortex superfluid. The intersting thing
happens when there exists
a small number delocalized and coherent defects (vacancies or interstitials) in the vortex crystal. These can
destroy the superfluidity but while preserving the underlying vortex crystal structure.
The resulting state can be regarded as a vortex supersolid, and it is equivalent to the CMI in our model.

\section{variational wavefunction for the CMI}
In order to go beyond mean field theory within a wavefunction approach, one can incorporate Jastrow factors
which build in interparticle correlations. Such Jastrow wavefunctions for an N-boson system take the form
\begin{equation}
\Psi(r_1,r_2,\ldots,r_N) = {\rm e}^{- \sum_{i,j} \tilde{v}(r_i-r_j)} \Psi_{MF} (r_1,r_2,\ldots,r_N)
\end{equation}
The simplest example of such a state is the Gutzwiller wavefunction for which $\tilde{v}(r_i-r_j)$ is nonzero and
positive for $r_i=r_j$ and is zero if $r_i \neq r_j$. Such a Gutzwiller factor builds in correlation effects by
suppressing those configurations in which multiple bosons occupy the same site. Such a Gutzwiller
wavefunction is however inadequate to describe superfluid and insulating states of bosons. The reason
is that such short range Jastrow factors do not correctly encode the small momentum behavior of the
equal time density structure factor. For a superfluid which supports a linearly dispersing sound mode
at low momentum,
we need to have the Fourier transformed Jastrow factor $v (q \to 0) \sim 1/q$, while a gapped
insulator in 1D should have $v (q \to 0) \sim 1/q^2$, as discussed in Ref.~26.
\begin{figure}[t]
\includegraphics[width=3.2in]{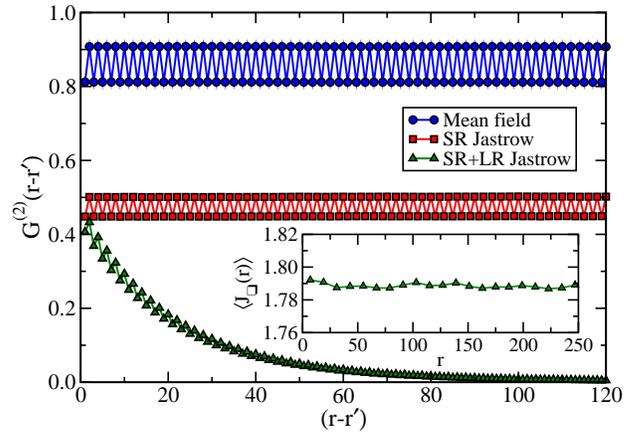}
\caption{Boson correlator
 $G^{(2)} (x-x')$
$= \langle a^\dagger_x  a_{x'} \rangle$ computed using variational Monte Carlo method
for: (i) the mean field CSF state (circles) with off-diagonal long
range order, (ii) a correlated
CSF with a Gutzwiller factor (squares), and (iii) the CMI state with a long-range
Jastrow (triangles) which leads to an exponential decay of $G^{(2)} (x-x')$.
The oscillations stem from low energy boson modes
at momenta $k=0$ and $k=\pi$. (Inset) Nonzero staggered current order
$(-1)^x \langle J_{\Box}(x) \rangle$ in the mean field CSF and in the CMI.}
\label{fig:varwfn}
\end{figure}

In order to describe the chiral Mott insulator, we therefore fix $\Psi_{MF}$ to be the mean field chiral
superfluid, and choose the Jastrow factor to have two pieces, a short range (on-site) Gutzwiller piece $v_{SR}$
and a long range part $v_{LR}$. We parametrize the on-site Gutzwiller factor to have strength $g_S$,
and choose the long range Jastrow factor to have a Fourier transform,
$v_{LR} (q) =1/(1 - \cos q)$, and with a strength $g_L$.
Using this parametrization, we have computed
the properties of the resulting variational wavefunction using standard
Monte Carlo sampling, averaging over $10^6$-$10^7$ boson configurations on the two-leg ladder for
system sizes upto $L=500$.

Fig.~\ref{fig:varwfn} shows the computed 2-point correlation function $G^{(2)}(x-x') = \langle a^\dagger_{x} a^\pdg_{x'} \rangle$
for different choices of $g_S$ and $g_L$. For $g_S=g_L=0$, we find the mean field result where $G^{(2)}(r)$
saturates to a nonzero value, while exhibiting oscillations which correspond to the fact that there are boson
modes at both momenta $k=0$ and $k=\pi$ making up the condensate wavefunction. For $g_S=1$ and $g_L=0$,
the local Gutzwiller factor suppresses multiple occupancy - this weakens but {\it does not} destroy
the off-diagonal long range order present in the mean field state. For $g_S=1$ and $g_L=0.1$, we find a rapid
exponential decay of $G^{(2)}(x-x')$ (shown in Fig.~8), with a correlation length $\xi \sim 25$ lattice spacings obtained from a fit
to the numerical data. This state thus has
exponentially decaying boson correlations indicating a gapped ground state.

\begin{figure}[t]
\includegraphics[width=3.4in]{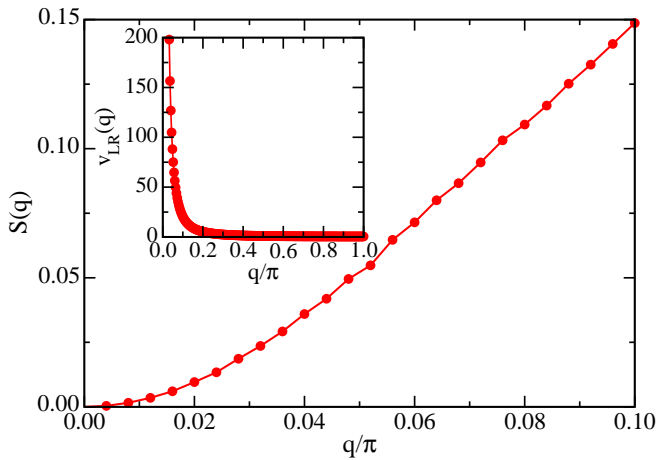}
\caption{Behavior of the density structure factor $S(q)$ in the chiral Mott state with both short and
long range Jastrow factors, showing $S(q) \sim q^2$ at small momenta. Inset depicts the singular
long range Jastrow factor $v_{LR} \sim 1/q^2$ which is necessary to obtain this insulating phase.}
\label{Fig:jastrow}
\end{figure}

This is in agreement with the behavior of the boson density structure factor $S(q)$. For $g_S=1$ and
$g_L=0.1$, we find, as shown in Fig.~\ref{Fig:jastrow},
that the structure factor exhibits a $\sim q^2$ behavior at small momentum which
suggests, within the Feynman single-mode approximation, a gapped ground state. From the inset of Fig.~\ref{Fig:jastrow},
we find that this state nevertheless has an average nonzero staggered current in each plaquette which is inherited
from the mean field state, although it is slightly suppressed due to the on-site Gutzwiller correlations. This
Jastrow correlated state thus provides a simple variational ansatz for the Chiral Mott insulator ground state.
\vspace{1cm}

\section{Experimental detection}
How might the CMI be realized in experiments and detected? One possibility would to try to realise it
in a Josephson junction ladder with a flux of $hc/4e$ per plaquette. The detection of this phase would then
involve a measurement of the resistivity to verify that it is an insulator along with a measurement of local
magnetic moments to detect the pattern of staggered moments generated by the circulating currents.
An estimate shows that junctions with a coupling of about 1 Kelvin and lattice parameter of 10 $\mu$m could
produce staggered fields of about 1 nT resulting from the staggered moments, which can be measured by high precision
Supercondcuting Quantum Interference Device (SQUID) microscopy.

Another possibility is to use synthetic magnetic fields to generate the $\pi$ flux in a system of
cold atoms in an optical lattice. The $\pi$ flux will then produce twin peaks in the single particle
momentum distribution, which will be sharp in the CSF and get broadened by the interaction in the CMI and MI.
There are two possible ways to distinguish between the CMI and the MI experimentally. 1) Intermode coherence can
be tested for by Bragg reinterference of the $k=0$ and $k=\pi$ peaks. \cite{bragg.epjd2005} The CMI will display
coherence between the two modes while the MI will not. 2) The presence of staggered current patterns can be detected
 directly using a quench experiment.~\cite{killi.pra2012} The staggered currents exist on bonds of a lattice circulating in
 a particular way (clockwise or anticlockwise) around the plaquettes. If the hopping on
 certain bonds (say those between the legs of the ladder)
 were to be suddenly turned off by a quench, the current loops would be broken resulting in charge accumulation and
depletion at lattice sites at later times, resulting in a time-dependent staggered modulation of the local charge density.
This density modulation can then
be used to back out the original pattern of currents flowing on the bonds of the lattice. Such modulations would also
be present in the CSF state, where the density modulations and oscillations would be likely to persist for longer, whereas
they would be short lived in the CMI due to heating effects from the quench-induced nonequilibrium currents.
The observation of such quench induced
density dynamics (albeit short-lived) together with a charge gap would be an unambiguous signature of an intermediate
CMI state.

\section{Generalizations}

We next discuss potential generalizations of our work along various directions -
although we do not have, at this stage, detailed numerical
studies on these examples, they are sufficiently well motivated physically and worth examining in more detail in
the future. We begin by noting the generalization to the case of staggered flux on the ladder, where we have
alternating fluxes $\Phi,-\Phi$ with $0 < \Phi < \pi$. We then discuss a
possible generalization of the FFBH model to 2D.

\subsection{Staggered flux on a 2-leg ladder}
Instead of a uniform $\pi$-flux per plaquette, let us imagine having staggered fluxes $\Phi,-\Phi$ on the two-leg
ladder, with $0 < \Phi < \pi$. The resulting band dispersion of
noninteracting particles exhibits a single non-degenerate minimum at $k=0$. Condensing into this minimum
leads to a superfluid with a unique current pattern, which alternates from plaquette to plaquette but whose
sense is entirely locked to the underlying flux pattern since the Hamiltonian itself breaks time-reversal
symmetry. In this sense, moving away from the fully frustrated limit towards a staggered flux state is like
turning on a ``magnetic field'' which couples to the underlying Ising order associated with loop currents.
Thus, the staggered loop currents survive to arbitrarily large $U/t$, in much the same way as the
paramagnetic phase of the usual Ising ferromagnet acquires a nonzero magnetization in a uniform
magnetic field at any finite temperature. Turning to the quantum phase diagram of the staggered flux
Bose Hubbard or quantum XY models,
we lose the distinction between the CMI and the MI, so the Ising transition associated
with this distinction will be replaced by a crossover. We thus expect to have a single
BKT transition from a superfluid to an insulator, with non-vanishing staggered loop currents at all values of
$U/t$. Numerical studies confirming this prediction would be valuable.

\subsection{2D analogue of the bosonic CMI state}

\begin{figure}[t]
\includegraphics[width=1.5in]{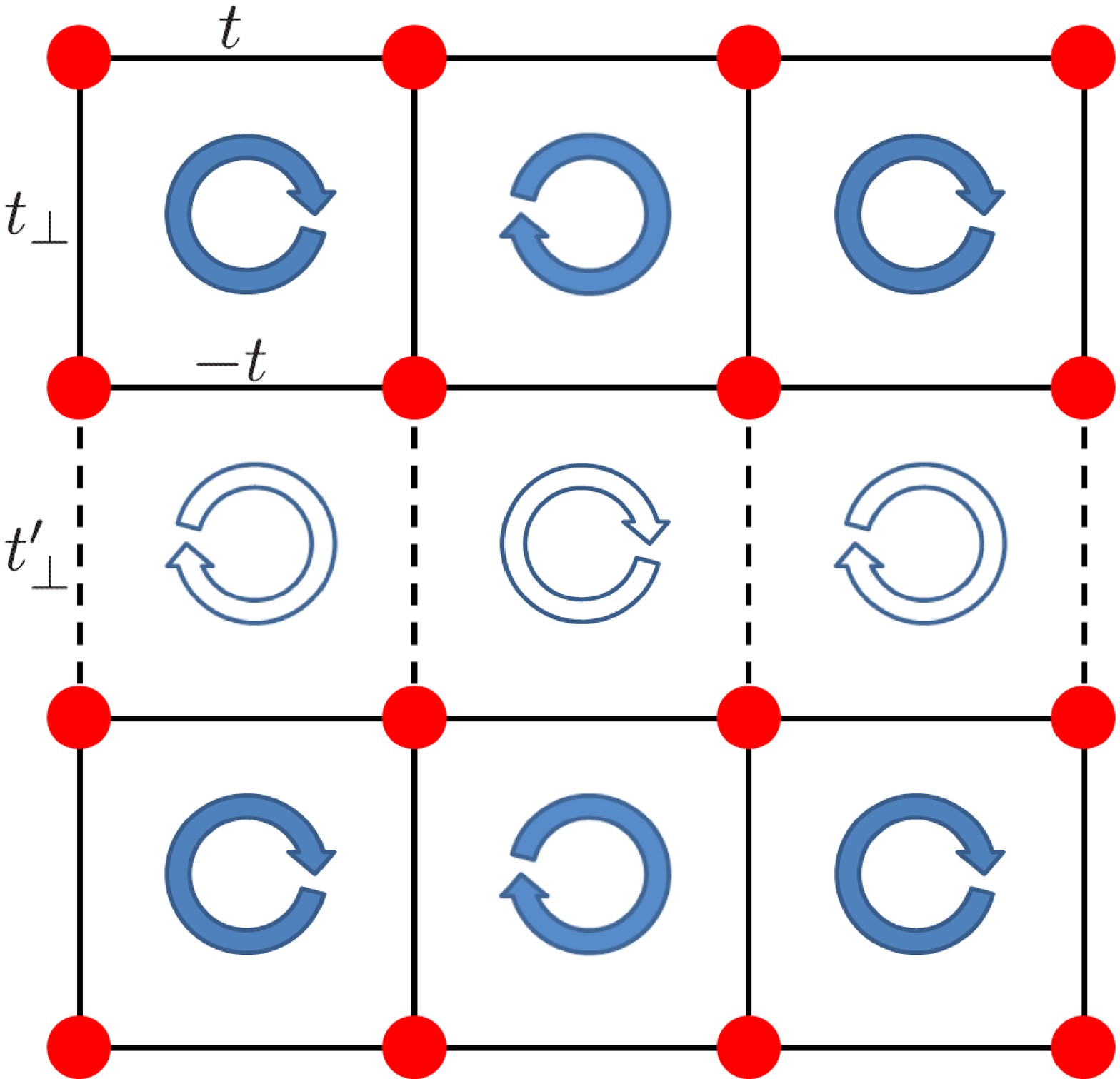}
\includegraphics[width=1.2in]{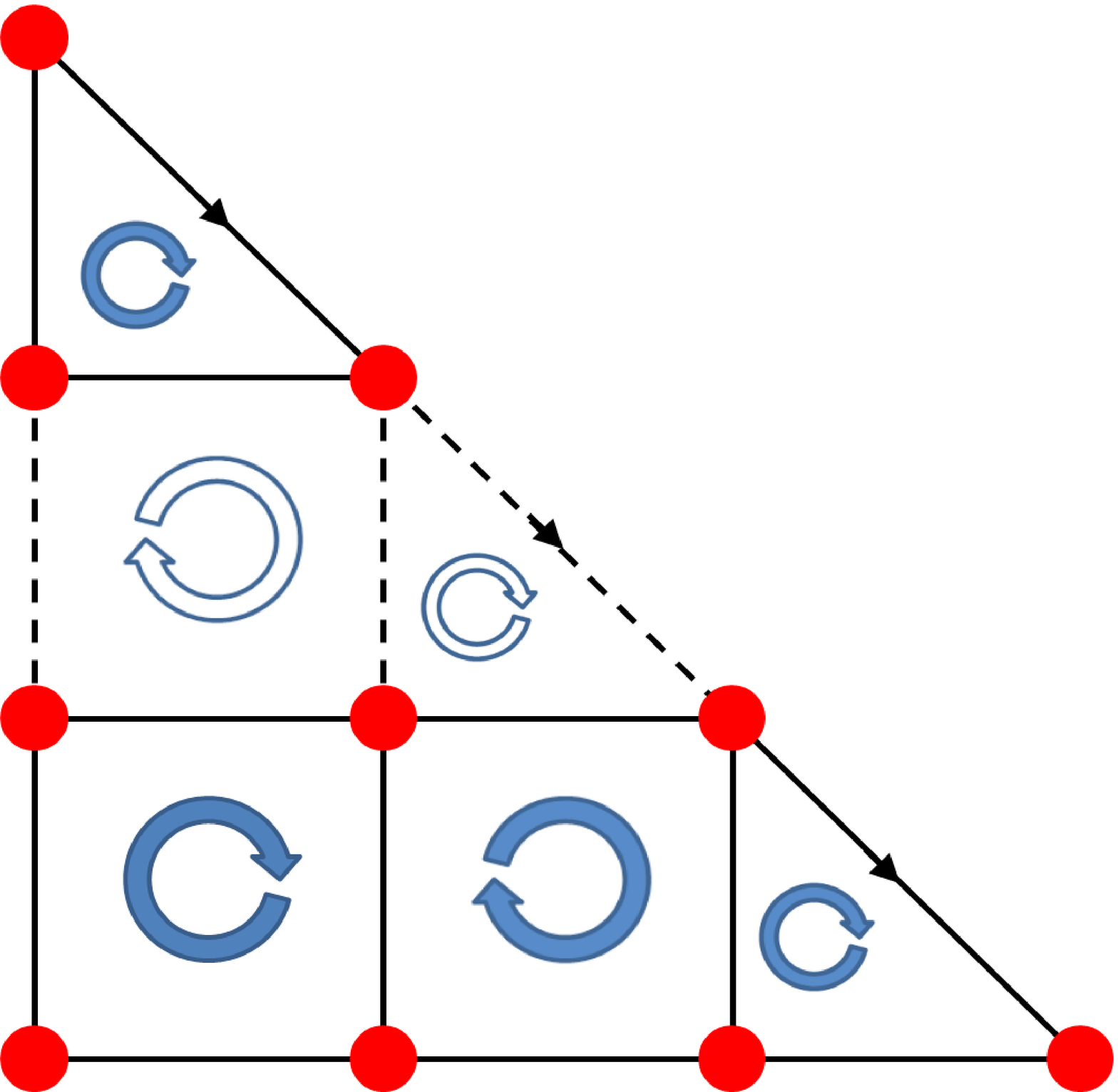}
\caption{Left panel: Sketch of a staggered loop current state in a 2D CMI state in a fully frustrated Bose
Hubbard model of weakly coupled two-leg ladders with $\pi$-flux per plaquette. Right panel: Sketch of
the [11] edge depicting edge current along such a 2D CMI state.}
\label{fig:2dcmi}
\end{figure}

The 2-leg ladder FFBH model we have studied here provides a nice example of an unusual weak
Mott insulator state which supports loop currents since the virtual boson number fluctuations permit
the bosons to `sense' the flux on each plaquette even though they are insulating at longer length
scales. Imagine weakly coupling many such ladders lying next to each other to form an anisotropic
version of the 2D FFBH model, where each square plaquette still has $\pi$-flux, but where the boson
hopping takes on intra-chain values $\pm t$ along alternate chains, with a modulated interchain
hopping which takes on a value $t_\perp$ within a ladder (intra-ladder rung hopping) and $t'_\perp$
between adjacent ladders (inter-ladder hopping). The sketch of this model is shown in Fig.~\ref{fig:2dcmi}
(left panel).
When $t'_\perp=0$, we get a set of decoupled FFBH ladders which support a fully gapped CMI state.
We thus expect that this phase will be stable so long as the gap in the CMI state is much larger than
the inter-ladder hopping $t'_\perp$.
The charge gap in the CMI would render this stable against Bose condensation into a 2D superfluid.
Similarly, the Ising gap associated with the broken $Z_2$ time-reversal symmetry  would render this 2D CMI
stable against disordering of the current loop order for small values of $t'_\perp$. Based on the fact that
vortices repel each other, and would prefer a global $(\pi,\pi)$ antiferromagnetic ordering pattern in 2D,
we expect the current loops on the different ladders to phase lock in-step producing a fully gapped 2D
CMI state with staggered loop currents as shown in the left panel of Fig.~\ref{fig:2dcmi}.
Such a CMI state would lead to a chiral edge current along any
of the [11] edges of the lattice as shown in the right panel of Fig.~\ref{fig:2dcmi}.
Numerical studies of weakly coupled ladders within a classical XY model
calculation would be valuable to explore the full phase diagram as a function of $U/t$, $t_\perp/t$, and $t'_\perp/t$.

\section{Summary}
We have computed the accurate phase diagram of the FFBH model using
two different numerical techniques: Monte Carlo simulations and  the FS-DMRG method. We
have consistently obtained qualitatively similar phase diagrams depicting three different quantum
phases: CSF, CMI and MI. The CMI
phase is sandwiched between the CSF and the MI phase, and is a remarkable state of bosonic matter,
being an insulator but with staggered loop currents which spontaneously break time-reversal symmetry.
We have confirmed that the transition from CSF-CMI is BKT type and the transition from CMI to MI is of Ising type.
We have also shown that the boson momentum distribution has different properties in these phases.
We present the two different
physical pictures for the
CMI as an exciton condensate and the vortex supersolid. A variational wavefunction which can explain the CMI phase is
discussed. Finally, we have proposed possible
experimental signature of different quantum phases which can be realizable in the experiments involving cold atoms in optical lattices and Josephson junction
arrays, and discussed higher dimensional realizations of the CMI in a FFBH model.

\section{Acknowledgments}
We would like to acknowledge useful discussions with B. P. Das and
support from the Department of Science and Technology, Government of
India (SM and RVP), CSIR, Government of India (RVP) and NSERC of
Canada (AP). We would also like to acknowledge GARUDA, India for computational facilities.

\appendix
\section{Derivation of classical XY model}
We define
the new rotor
variables to satisfy the following commutators:
\bea
\left[{\rm e}^{-i\varphi^\alpha_{x}}, {\cal L}^\beta_{x'} \right] &=& {\rm e}^{-i\varphi^\alpha_{x}} \delta^{\alpha\beta} \delta_{x,x'} \\
\left[{\rm e}^{-i\varphi^\alpha_{x}}, {\rm e}^{-i\varphi^\beta_{x'}} \right] &=& 0 \\
\left[{\cal L}^\alpha_{x} , {\cal L}^\beta_{x'} \right] &=& 0
\eea
where $\alpha/\beta = a,b$. In terms of these, the Hamiltonian takes the form
\bea
H_{\rm rotor} &=& - 2 \tt \sum_x \cos(\varphi^a_{x} - \varphi^a_{x+1})  + 2 \tt \sum_x \cos(\varphi^b_{x} - \varphi^b_{x+1}) \nonumber \\
&-& 2 \tt_\perp
\sum_x\cos(\varphi^a_{x} - \varphi^b_{x}) \nonumber\\
&+& \frac{U}{2} \sum_x \left[({\cal L}^{a}_x)^2  + ({\cal L}^{b}_x)^2)\right] - \mu \sum_x ({\cal L}^{a}_x + {\cal L}^{b}_x),
\label{Hrotor}
\eea
where we have allowed the rotor hopping to be proportional to the original boson hopping, with $\tt_\perp/\tt = t_\perp/t$.
We expect the proportionality constant to be such that
$\tt = t \psi^2 \approx t \rho$ where $\rho$ is the boson density per site.

At $\mu=0$, which corresponds to an average of one boson per site,
we can easily go to the path integral representation.
We start with the partition function
$
Z = {\rm Tr} ({\rm e}^{-\beta H_{\rm rotor}})
$
and use Trotter discretization to set
$\exp(-\beta H_{\rm rotor}) = [\exp (-\epsilon H_{\rm rotor})]^{L_\tau}$, with $\epsilon = \beta/L_\tau$. The
partition function then takes the form
\bea
Z = \sum_{\{\{\varphi_{x,\tau}\}\}} \la \{\varphi_{x,0}\} | {\rm e}^{-\epsilon H_{\rm rotor}} | \{\varphi_{x,L_\tau-1}\} \ra \ldots \nonumber \\
\ldots \la \{\varphi_{x,1}\} | {\rm e}^{-\epsilon H_{\rm rotor}} | \{\varphi_{x,0}\} \ra
\eea
where we have chosen to compute the trace in the angle basis and intermediate states are also in this basis.
Here, single curly brackets refer to an angle-configuration at a fixed `time'-slice, while double curly brackets
denote angle-configurations over all of space-`time'.
Let us consider the matrix element
\be
\la \{\varphi_{x,\tau}\} | {\rm e}^{-\epsilon H_{\rm rotor}} | \{\varphi_{x,\tau-1}\} \ra.
\ee
Since $\epsilon \to 0$, we can set
\begin{multline}
\la \{\varphi_{x,\tau}\} | {\rm e}^{-\epsilon H_{\rm rotor}} | \{\varphi_{x,\tau-1}\} \ra  \\
\approx \la \{\varphi_{x,\tau}\} | {\rm e}^{-\epsilon V_{\rm rotor}} {\rm e}^{-\epsilon K_{\rm rotor}} | \{\varphi_{x,\tau-1}\} \ra  \\
\approx {\rm e}^{-\epsilon V_{\rm rotor}(\{\varphi_{x,\tau}\})}
\la \{\varphi_{x,\tau}\} |  {\rm e}^{-\epsilon K_{\rm rotor}} | \{\varphi_{x,\tau-1}\} \ra
\end{multline}
where $V$ and $K$ are the potential ($\varphi$-diagonal) and kinetic ($\varphi$-off-diagonal) terms in
the rotor Hamiltonian. Explicitly,
\bea
V_{\rm rotor}(\{\varphi_{x,\tau}\}) &=& - 2 \tt \sum_x \cos(\varphi^a_{x,\tau} - \varphi^a_{x+1,\tau}) \nonumber \\
& +& 2 \tt \sum_x \cos(\varphi^b_{x,\tau} - \varphi^b_{x+1,\tau}) \nonumber \\
&-& 2 \tt_\perp \sum_x\cos(\varphi^a_{x,\tau} - \varphi^b_{x,\tau}).
\eea
and
\be
K_{\rm rotor} = -\frac{1}{\epsilon^2 U}
\sum_{x,\tau} [\cos(\varphi^a_{x,\tau} - \varphi^a_{x,\tau+1}) + \cos(\varphi^b_{x,\tau} - \varphi^b_{x,\tau+1})]
\ee
This gives us the partition function in terms of a classical action in one higher dimension
\be
Z = \sum_{\{\{\varphi_{x,\tau}\}\}} {\rm e}^{-S^{1+1}_{\rm cl}[\varphi]}
\ee
where
\begin{widetext}
\bea
S^{1+1}_{\rm cl} &=&  - \sum_{x \tau} \left[J_\parallel \cos(\varphi^a_{x+1,\tau} - \varphi^a_{x,\tau}) -
J_\parallel \cos(\varphi^b_{x+1,\tau} -
\varphi^b_{x,\tau}) + J_\perp \cos(\varphi^a_{x,\tau} - \varphi^b_{x,\tau}) \right] \nonumber \\
&-& J_\tau \sum_{x \tau}
\left[ \cos(\varphi^a_{x,\tau+1} - \varphi^a_{x,\tau}) + \cos(\varphi^b_{x,\tau+1} - \varphi^b_{x,\tau}) \right],
\eea
\end{widetext}
with $2 \epsilon \tt=J_\parallel$, $2 \epsilon \tt_\perp = J_\perp$, and $1/\epsilon U = J_\tau$. We see that this has the form of an anisotropic $XY$ model.
In order to get the properties of the quantum model at a fixed inverse temperature $\beta \tt$, we must take
the `time'-continuum limit of $\epsilon \to 0$, sending $J_\parallel \to 0$, $J_\perp \to 0$, and $J_\tau \to \infty$, while
keeping fixed $J_\perp/J_\parallel = \tt_\perp/\tt$  and $J_\tau J_\parallel = 2 \tt/U$. The inverse temperature
$\beta \tt$ is then given by $\epsilon \tt L_\tau$ and thus depends on the chosen value of $\epsilon \tt$ (which
must be taken to be very small) and the
size of the simulation cell in the `time'-direction. We set $\epsilon = 1/\sqrt{2 U \tilde{t}}$ which
leads to $J_\parallel = J_\tau = \sqrt{2 \tilde{t}/U}$.

\section{Momentum Distribution}

The main result of this appendix is to show that the peaks in the CSF phase of the
ladder model have a non-universal power law divergences, with equal exponents, at $k=0,\pi$.
This is consistent with our DMRG simulation results of the FFBH model
which show two such equal height and power
law divergent peaks in momentum distribution in the CSF state (shown in Fig.~\ref{Fig:nk} and discussed
in Sec. V).

To analytically calculate the momentum distribution in the CSF on the ladder, we begin by carrying
out a classical minimization of the rotor potential energy in Eq.~\ref{Hrotor}, which
leads to the equations
\begin{widetext}
\bea
\tt [\sin (\varphi^a_x - \varphi^a_{x+1}) +  \sin (\varphi^a_x - \varphi^a_{x-1})] + \tt_\perp \sin (\varphi^a_x - \varphi^b_{x}) &=& 0, \\
\tt [\sin (\varphi^b_x - \varphi^b_{x+1}) +  \sin (\varphi^b_x - \varphi^b_{x-1})] + \tt_\perp \sin (\varphi^a_x - \varphi^b_{x}) &=& 0.
\eea
\end{widetext}
It is straightforward to check that we satisfy these equations by
substituting $\cos \varphi^a_x = u_0$, $\cos \varphi^b_x = v_0$, $\sin \varphi^a_x = \pm v_0 (-1)^x$, and
$\sin \varphi^b_x = \pm u_0 (-1)^x$, where $u_0,v_0$ are defined via Eqns.\ref{equk},\ref{eqvk}.
For now, let us stick with the `+' solution.
We can write this solution as $\varphi^a_x = (-1)^x \Psi$, $\varphi^b_x =  (-1)^x (\pi/2-\Psi)$ where $\cos \Psi = u_0$
and $\sin\Psi = v_0$. To take small fluctuations into account, we must set
\bea
\varphi^a_x &=& (-1)^x \Psi + \tphi^a_x, \\
\varphi^b_x &=& (-1)^x (\pi/2-\Psi) + \tphi^b_x.
\eea
Using the above minimization condition, we find the Hamiltonian for small fluctuations,
\begin{widetext}
\bea
H^{\rm small}_{\rm rotor} &=& \tt \cos(2\Psi) \sum_x \left[
(\tphi^a_{x} \!-\! \tphi^a_{x+1})^2  + (\tphi^b_{x} \!-\! \tphi^b_{x+1})^2\right] + \tt_\perp \sin(2\Psi)  \sum_x (\tphi^a_{x} \!-\! \tphi^b_{x})^2 \nonumber\\
&+& \frac{U}{2} \sum_x \left[({\cal L}^{a}_x)^2  + ({\cal L}^{b}_x)^2)\right] - \mu \sum_x ({\cal L}^{a}_x + {\cal L}^{b}_x)
\eea
\end{widetext}
Further, for small fluctuations, we must set $\exp(-i \tphi) \approx 1 - i\tphi$, so that
the full commutation relations are replaced  by
\bea
\left[\tphi^\alpha_{x}, {\cal L}^\beta_{x'} \right] &=& i \delta^{\alpha\beta} \delta_{x,x'} \\
\left[\tphi^\alpha_{x}, \tphi^\beta_{x'} \right] &=& 0 \\
\left[{\cal L}^\alpha_{x} , {\cal L}^\beta_{x'} \right] &=& 0
\eea
so that $\tphi$ and ${\cal L}$ have exactly the same commutation relations as position/momentum variables.
Let us now define a shifted angular momentum
$\tilde{\cal L}^\alpha_x = {\cal L}^\alpha_x - \mu/U$ to absorb the chemical potential term,
which leads to the Hamiltonian (upto an additive constant we will ignore),
\bea
H^{\rm small}_{\rm rotor} &=& \tt \cos(2\Psi) \sum_x \left[
(\tphi^a_{x} \!-\! \tphi^a_{x+1})^2  + (\tphi^b_{x} \!-\! \tphi^b_{x+1})^2\right] \nonumber \\
&+& \tt_\perp \sin(2\Psi)  \sum_x (\tphi^a_{x} \!-\! \tphi^b_{x})^2 \nonumber\\
&+& \frac{U}{2} \sum_x \left[(\tilde {\cal L}^{a}_x)^2  + (\tilde {\cal L}^{b}_x)^2)\right]
\eea
Fourier transforming, we find
\begin{widetext}
\bea
H^{\rm small}_{\rm rotor} &=& \frac{1}{2} \sum_{q,\alpha=a,b}
 \left[2 \tt \cos(2\Psi)  (2\!-\!2\cos q)  \!+\! 2 \tt_\perp \sin(2\Psi)\right]  \tphi^\alpha_{q} \tphi^\alpha_{-q} \nonumber \\
\!&-&\! \frac{1}{2} \sum_q 2 \tt_\perp \sin(2\Psi)   (\tphi^a_{q} \tphi^b_{-q} \!+\! \tphi^a_{-q} \tphi^b_{q})
+ \frac{U}{2} \sum_{q,\alpha} \tilde {\cal L}^{\alpha}_q  \tilde {\cal L}^{\alpha}_{-q}.
\eea
\end{widetext}
This is like a problem of two coupled oscillators for each $q$, with each oscillator having a `mass' $=1/U$, but
with a $2\times 2$ spring constant matrix having eigenvalues
\bea
K_+(q) &=& 2 \tt \cos(2\Psi)  (2\!-\!2\cos q) \\
K_-(q) &=& 2 \tt \cos(2\Psi)  (2\!-\!2\cos q) + 4 \tt_\perp \sin(2\Psi).
\eea
This leads to a collective mode spectrum with two branches,
$
\omega_\pm  = \sqrt{U K_\pm(q)},
$
where $\omega_+(q) = c q$ for small $q$ behaves like an `acoustic phonon' mode, while $\omega_-(q)$ is like
a gapped `optical phonon' mode. Using the fact that $\sin (2\Psi) = \tt_\perp/\sqrt{\tt_\perp^2 + 4 \tt^2}$ and
$\cos (2\Psi) = 2 \tt /\sqrt{\tt_\perp^2 + 4 \tt^2}$, we find that
the `speed of sound' is
\be
c = \sqrt{\frac{4 U \tt^2}{\sqrt{4 \tt^2 + \tt_\perp^2}}} \approx \sqrt{\frac{4 U \rho t^2}{\sqrt{4 t^2 + t_\perp^2}}}
\ee
while the minimum gap to the `optical phonon' mode is at $q=0$ and given by
\be
\omega_-(q=0) = \sqrt{\frac{4 U \tt_\perp^2}{\sqrt{ 4 \tt^2 + \tt_\perp^2 }}} \approx \sqrt{\frac{4 U \rho t_\perp^2}{\sqrt{ 4 t^2 +
t_\perp^2 }}}
 \ee
These should be viewed as analogs of a Bogoliubov theory result for a  1D system with no Bose condensate
(i.e., no ODLRO).
The off-diagonal one-body density matrix may be computed by noting that
\be
\la a^\dg_x a^\pdg_y\ra \sim \rho {\rm e}^{i \Psi \Delta_-(x,y)}  \la {\rm e}^{i (\tphi^a_x - \tphi^a_y)} \ra
\ee
where $\Delta_\pm (x,y) = (-1)^x \pm (-1)^y$.
Within
the Gaussian theory, we get
\be
\la a^\dg_x a^\pdg_y\ra \sim \rho {\rm e}^{i \Psi \Delta_-(x,y)} {\rm e}^{-\frac{1}{2}\la (\tphi^a_x - \tphi^a_y)^2 \ra}
\ee
The expectation value in the exponential can be evaluated as
\be
\la (\tphi^a_x - \tphi^a_y)^2 \ra = \int_{-\pi}^{+\pi} \!\!\frac{d q}{2\pi}~\la \tphi^a_q \tphi^a_{-q} \ra \left[2 - 2 \cos(q (x-y)) \right]
\ee
We find
\bea
\la \tphi^a_q \tphi^a_{-q} \ra &=& \la \tphi^b_q \tphi^b_{-q} \ra = \frac{1}{4} \left(\sqrt{\frac{U}{K_+(q)}} + \sqrt{\frac{U}{K_-(q)}}\right)\\
\la \tphi^a_q \tphi^b_{-q} \ra &=& \la \tphi^b_q \tphi^a_{-q} \ra = \frac{1}{4} \left(\sqrt{\frac{U}{K_+(q)}} - \sqrt{\frac{U}{K_-(q)}}\right)
\eea
Since $K_+(q \to 0) \sim q^2 $, while $K_-(q \to 0)$ remains nonzero, the long wavelength (small $q$)
limit of all such correlations are given by
\be
\la \tphi^\alpha_q \tphi^\beta_{-q} \ra_{_{q \to 0}} \approx \frac{1}{4} \sqrt{\frac{U}{K_+(q)}}
\ee
Using this, and defining
\be
F(x-y) = \left(\frac{1}{|x-y|}\right)^{-\frac{1}{4\pi} \sqrt{\frac{U}{2 \tt \cos 2\Psi}}},
\ee
we find the long distance behavior of the one-body off-diagonal density matrix elements are given by
\bea
\la a^\dg_x a^\pdg_y\ra &\sim& \rho{\rm e}^{i \Psi  \Delta_-(x,y)} F(x-y) \\
\la b^\dg_x b^\pdg_y\ra &\sim& \rho {\rm e}^{- i \Psi \Delta_-(x,y) + i \frac{\pi}{2} \Delta_-(x,y)} F(x-y) \\
\la a^\dg_x b^\pdg_y\ra &\sim&\rho {\rm e}^{i \Psi \Delta_+(x,y) - i\frac{\pi}{2}(-1)^y} F(x-y) \\
\la b^\dg_x a^\pdg_y\ra &\sim& \rho {\rm e}^{-i \Psi \Delta_+(x,y) + i\frac{\pi}{2}(-1)^x} F(x-y)
\eea
Let us expand
\bea
\la a^\dg_x a^\pdg_y\ra &\sim &  \rho (\cos^2\Psi \!+\! \sin^2\Psi (-1)^{x-y} \nonumber \\
&+& i \Delta_-(x,y) \sin 2 \Psi ) F(x-y)
\eea
using which,
\bea
n_{aa}(k) &\equiv & \frac{1}{L} \sum_{x,y} \la a^\dg_x a^\pdg_y\ra {\rm e}^{i k (x-y)} \sim \rho \cos^2\Psi \sum_r F(r) {\rm e}^{i k r}\nonumber \\
&+& \rho \sin^2\Psi \sum_r F(r) {\rm e}^{-i \pi r}  {\rm e}^{i k r},
\eea
where $L$ is the linear system size. We find
$n_{aa}(k \to 0) \sim (\rho \cos^2\Psi) |k|^{-\alpha}$ and $n_{aa}(k \to \pi) \sim (\rho  \sin^2\Psi) |k-\pi|^{-\alpha}$, where
\be
\alpha = 1 - \frac{1}{4\pi} \sqrt{\frac{U}{2 \tt \cos 2\Psi}} \approx 1 - \frac{1}{4\pi} \sqrt{\frac{U \sqrt{4 t^2 + t_\perp^2}}{4 \rho t^2}}
\ee
Similarly, $n_{bb}(k \to 0) \sim ( \rho\sin^2\Psi) |k|^{-\alpha}$ and $n_{bb}(k \to \pi) \sim ( \rho\cos^2\Psi) |k-\pi|^{-\alpha}$, so that
$n_{aa}(k) + n_{bb})(k) \sim \rho ( |k|^{-\alpha} + |k-\pi|^{-\alpha})$.
We can also calculate cross correlators such as $n_{ab}(k) = \la a^\dg_k b^\pdg_k\ra$ and $n_{ba}(k)
= \la b^\dg_k a^\pdg_k \ra$. We find that $n_{ab}(k) = n_{ba}(k)$,
and that
$n_{ab}(k \to 0) \sim ( \rho\sin \Psi \cos\Psi) |k|^{-\alpha}$, while $n_{ab}(k \to \pi) \sim ( \rho\sin \Psi \cos\Psi) |k-\pi|^{-\alpha}$.

We have thus shown that the peaks in the CSF phase of the
ladder model have a non-universal power law divergences, with equal exponents, at $k=0,\pi$.
Similarly, two (broad) equivalent peaks in the momentum distribution also appear in the MI, but they need a
strong coupling expansion to compute as has been done, together with a collective mode analysis,
for the 2D FFBH model~\cite{sengupta.epl2010}.
At the BKT transition from the CSF to CMI, an analysis of which requires going beyond our
harmonic theory, we expect the exponent to take on a universal value $\alpha=3/4$. Such a bosonization
analysis will be described elsewhere.

\end{document}